论 文　　专题名称

# 流动手性与结构的几何和拓扑图像描述

邹文楠[①], 朱建州[②]*, 刘鑫[③]

① 南昌大学高等研究院，南昌 邮编：330031
② 南京市高淳区速诚基础与交叉科学研究中心，南京 邮编：211316;
③ 北京工业大学理论物理所，北京 邮编：10080;
*联系人, E-mail: jz@sccfis.org

**摘要**　　本文聚焦流动手性与结构相关问题，介绍通过几何与拓扑语言描述的最新甚至独特的理论研究结果。考虑到我们国内流体力学的教学和基础前沿研究情况，希望能够低起点高坡度地适宜从学生到教学与研究人员的更普遍读者，也力图阐明相关基本概念，梳理经典流动理论相关成果的微分形式表达，并阐述我们最近在拓扑流体动力学和湍流理论方面的探索。

**关键词**　　基础流体力学，湍流理论，拓扑，微分几何，螺（旋）度

## 0　引言

流体动力学如何描述才是根本和方便的？这是个教（授）、学（习）和研（究）中必须注意的问题；甚至，一门学科是否能得到突破也取决于描述语言是否合适。我们根据自身研究的体会和对我国现状的观察，以及最近在与其它一些国内一线教学与基础研究岗位上的同事交流中（包括关于一些非常具体的理论问题的讨论），特别地有如下的感受和共鸣：力学或一般物理学发轫于几何直观，为成为一门学问的初步系统发展所必需的严格推演而解析化、抽象化；但是，无论是学者的深刻理解还是学科的进一步发展，都必须回归几何直观，结合分析，进行严格系统的几何化。这里"严格"只代表"准确"，不应晦涩，而应鲜活。这样的理解和观念有必要在教学和研究的设计和布局中得到吸收和重视。这本是一个自然的过程，但是，通过在国内外流体力学基础教育和前沿研究方面的观察和对比，尤其在长久以来湍流问题不能妥善解决的压力下，我们觉得反而应该稳下心来，注重基础，从根本入手，并把这样的想法落到入门（教学）到前沿（研究）的实处，对今后流体力学学科的稳固和取得实质进展都有重要的意义。

现代意义上的微分形式，以及以楔积和外微分结

构形成外代数的想法，是由著名法国数学家埃里·卡当(Elie Cartan)在 1899 年引入的[1]。兼具本体论（如黎曼几何之于弯曲时空）和方法论（描述一般流形）的意义。Arnold[2] 开创用其对理想流动的新颖基础描述，并获拥簇和蓬勃发展。比如，最近有将流动统计理论几何化的工作[3] 。但事实上相关的思想和语言在流体力学早期的发展中就已经得到一定的理解和应用。比如，最近文献[4] 就引用吉布斯的报告指出"在矢量获得通用的一百多年前，达朗贝尔、欧拉和拉格朗日等人的流体力学著作中就已经普遍使用微分 1-形式"。

这样表述的物理定律具有特别简洁的形式，而且计算是自然和自动的，不受空间维数的限制，也不必记住矢量分析中的许多公式，这意味着它们或许也是流体力学新思想的重要源泉。微分形式表达的物理量是这样的一些量，在它们上面可以施行外积即反对称化的直积运算、外微分与微分算子的外积运算以及积分运算等。微分形式表述的物理定律可以说是集微分形式和积分形式为一体的。

近半个多世纪以来，流体力学的几何化其实已经比较系统，出现了大量的研究成果，甚至一些专著，但是似乎还不够"流行"。研究文献上看，国内用这种基本语言进行基础讨论的寥寥无几，缺乏交流群体和氛围。

流体力学的应用，弥补在解决复杂问题方面的薄弱，归根结底是要加强基础。中国在过去半个多世纪由于各种原因，总体上基础方面缺乏足够的引导，有所贻误。也许，只有在基础的入门（教学）和前沿（研究）中回归根本，足够重视和落实，才可能后发先至。

本文有三个目的：一个是希望通过介绍一些基本概念和最新研究来传播相关思想，引起重视；二是希望整理和梳理相关的内容，提出前进的思路和观点；三是探索突破前沿研究问题（比如，湍流）的想法，并希望也能对相关教学有所启发。内容安排如下：此后的第一节至第四节是一些基本的拓扑与几何背景概念，和经典流动的微分几何表述。已经熟悉这些背景知识的读者可跳过，直接进入后面三位作者有分有合，各自主要负责撰写的章节；最后一节我们给出总结性讨论和展望。下面简介三个核心内容章节：

第五节主要介绍朱建州在经典流体动力学与局部螺旋度（local helicity/spirality）相关的最近工作，以及相关湍流研究的讨论。这部分内容用几何的语言描述，其实是经典流体力学的延伸发展，加入了作者的一些湍流方面的思考和尝试（比如，非经典的 --- 可能是不唯一、随机的 --- 虚拟"冻结"或 Lie 携带速度，以及湍流的局部几何对象不变约束，等等。）

第六节结合刘鑫最近工作系统展开介绍（全局）螺度相关的拓扑学知识，尤其是纽结理论，也间或点到朱建州最近关于有螺湍流统计动力学的工作。这一节介绍的螺度及其守恒是流体工作者比较熟悉和比较容易理解的，即便有螺湍流统计动力学的基础研究，虽然并非大众化的，但总归还是比较属于流体物理或湍流物理工作者"圈中的"；拓扑流体力学也是经典流体的"正常"延伸发展。但是涉及与现代理论物理和数学纽结相关的先进内容的交叉融合，对通常流体学者的透彻理解可能是高坡度上升的过程，通读它是可以有益的。

第七节主要介绍邹文楠独创的不同于经典流体力学的一套湍流唯像规范场论。

第八节总结和进一步讨论螺度在工程力热声相关应用和基础之间密切且直接的联系。

# 1 微分流形：概念基础及其对流动理论的意义

用 $\Re$ 表示实数域、$\mathcal{E}^n$ 表示全体有序的 $n$ 个实数所组成的数组的集合，即

$$\mathcal{E}^n = \{x = (x_1, \ldots, x_n) | x_i \in \Re, 1 \leq i \leq n\}, \quad (1)$$

实数 $x_i$ 称为点 $x \in \mathcal{E}^n$ 的第 $i$ 个坐标，赋予空间 $\Re^n$ 线性构造和距离函数使之成为 $n$ 维矢量空间，亦即是 $n$ 维欧式空间[1]。

设 $f$ 为两个欧氏空间之间的映射，$f: \mathcal{E}^n \to \mathcal{E}^n$。欧氏空间（连续统）的连续性定义为：$f$ 在 $x \in \mathcal{E}^n$ 连续，假如给定 $\varepsilon > 0$，存在 $\delta > 0$，使得当 $|y - x| < \delta$ 时有 $|f(y) - f(x)| < \varepsilon$。若 $f$ 在每点 $x \in \mathcal{E}^n$ 满足这个条件，则称 f 为连续映射。$\mathcal{E}^n$ 的子集 N 称为点 $p \in \mathcal{E}^n$ 的一个邻域，假如对于某个实数 $r > 0$，以 $p$ 为中心、$r$ 为半径的闭球包含在其内。由连续性定义，假如对于任何

$x \in \mathcal{E}^n$，以及$f(x)$在$\mathcal{E}^n$内的邻域，$f^{-1}(x)$为$x$在$\mathcal{E}^n$内的一个邻域[5]。

在空间的每点定义一组邻域，再从邻域引出连续映射的适当定义，但摒除距离的概念，这可以成为拓扑空间的出发点。一般来说，拓扑（结构）是指这样的集族σ，它满足两个条件：σ的任何两个元的交和σ的任一子族的元的并仍为σ的元。显然，空集∅和集合 X = ∪{U: U ∈ σ}必须是σ的元，且σ的每一个元都是 X 的子集。于是，称 X 为拓扑σ的空间，而σ是关于 X 的一个拓扑，并称(X,σ)为拓扑空间（当不会引起误解时，简说成"X 为拓扑空间"）。通常意义上拓扑σ的元叫做关于σ的开集或σ-开集，如果只涉及一种拓扑，就简称为开集。拓扑空间的拓扑也可以借助点的邻域来描述，故又称为具有邻域拓扑的空间。Hausdorff 给出了拓扑空间的如下定义：

定义 0.1 拓扑空间定义为一个集合 X，连带 X 的每个点 $x$ 有一簇子集$U_x$，称为邻域，这些邻域满足：（1）每个元素 $x$ 至少有一个邻域，$x$ 在它的每个邻域里；（2）$x$ 的任何两个邻域的交集是它的一个邻域；（3）如果 $y$ 是$U_x$的一个点，则存在一个$U_y \subseteq U_x$；（4）如果$x \neq y$，则存在$U_x$和$U_y$，使得$U_x \cap U_y = \emptyset$。

下面引进两条可数性公理（所谓可数集是每个元素能与自然数集 N 的每个元素之间能建立一一对应的集合）：（a）对于每一点 $x$，子集$U_x$所组成的集是可数集；（b）所有不同的邻域所组成的集是可数集。拓扑空间(X,σ)内的集 U 叫做点 $x$ 的邻域，当且仅当 U 包含有 $x$ 的一个开集（虽然一个点的邻域不必一定为开）[6][7]。

称点 $x$ 为拓扑空间(X,σ)的子集 A 的聚点（又称凝聚点或极限点），当且仅当 $x$ 的每个邻域包含异于 $x$ 的 A 中的点。拓扑空间的开集是它包含的所以点的邻域，拓扑空间的闭集包含它的所有聚点。此外，拓扑空间的子集为开集，当且仅当它的余集是闭集，反之亦然。一个空间或一个空间的子集是紧致的，如果每一个无穷子集都有聚点。如果一个集合不管怎样分成两个互不相交的子集，其中一个子集至少含有另一个子集的一个聚点，这个集合就称为是连通的。如果一个空间是它的一个可数子集的闭包（即子集本身加上它的聚点），它就是可分离的[7][6]。

现在说明什么是"连续映射"、什么是"同胚"。设 X 和 Y 是拓扑空间，映射$f: X \to Y$为连续，假如对于 X 的每个点 $x$，以及$f(x)$在 Y 内的任意邻域 N，集合$f^{-1}(x)$为 $x$ 在 X 内的邻域。映射$f: X \to Y$叫做一个同胚，假如它是一对一的连续满射，并且有连续的逆映射。如果这样一个映射存在，就称 X 同胚于 Y，或 X 拓扑等价于 Y。拓扑学研究几何图形的这样一些性质，它们在图形被弯曲、拉伸/压缩或任意的变形下保持不变，只要在变形的过程中既不使原来不同的点融为同一个点，又不使新的点产生。换句话说，这种变换的条件是：在原来图形的点与变换了的图形的点之间存在一个一一对应，并且邻近的点变成邻近的点，即满足连续性，这样的一个变换就叫做同胚（变换）[5][7]。

流形是欧氏空间的推广，是可以在其上做微积分的一种拓扑空间[8]。

定义 1.1 设 M 是 Hausdorff 空间，即其任两个不同点必包含不相交的邻域。若对任意一点$x \in M$，都有 $x$ 在 M 中的一个邻域 $U$ 同胚于 $n$ 维欧氏空间$\mathcal{E}^n$的一个开集，则称 M 是一个 $n$ 维流形（或称拓扑流形）。

定义 1.2 设 $M$ 是一个 $n$ 维流形，如果在 M 上给定了一个坐标卡（局部坐标系）集$\mathcal{B} = \left\{\left(U_i, \varphi_{U_i}\right)_{i \in I}\right\}$满足下列条件，则称它是 M 的一个 $C^r$ 微分结构：(1){$U_i, i \in I$}是 M 的一个开覆盖；(2)属于它的任意两个坐标卡是$C^r$相容的；(3)它是极大的，即对于 M 的任意一个坐标卡$(V, \varphi_V)$若与属于它的每一个坐标卡都是$C^r$相容的，则它自身必属于$\mathcal{B}$。

若在 M 上给定了一个 $C^r$ 微分结构，则称 M 是一个 $C^r$-微分（光滑）流形。

在流形上给定一个微分结构之后，在每一点的附近可以用线性空间来近似，即可以引进定义有加法和数乘的切空间和余切空间等概念。在点$p \in M$上的一个切矢量是全体在点 $p$ 有同一个切矢量的参数曲线的集合，$p$ 点上的所有切矢量组成切空间$TM_p$，流形上所有点的切空间（的笛卡尔积）构成流形 M 上的

切丛TM: $=\cup_{p\in M} TM_p$。$2n$ 维（底空间 $n$ 维加切空间 $n$ 维）的切丛具有自然的微分流形结构：自然投影 $\pi: TM \to M$ 将切空间的元素 $V \in TM_p$ 映射到点 $p \in M$，是光滑满映射；其逆映射 $\pi^{-1}: p \to TM_p$ 又称为切丛在 $p$ 点上的纤维。流形上的一个矢量场称为切丛的一个（横）截面，即矢量丛的一个截面在每点 $p \in M$ 赋予了该点切空间 $\pi^{-1}(p)$ 的一个矢量，而截面的加法和数乘将通过每个纤维来实现。

余切丛 $T^*M$ 通常作为切丛 TM 的对偶来引入，通过线性形式即微分 1-形式或余切矢量来构造，即对矢量场 $V \in TM$ 和微分 1-形式场 $\omega \in T^*M$，有

$$\omega: TM \to \Re | r = \langle V, \omega \rangle。 \quad (2)$$

点 $p \in M$ 的余切空间 $T^*M_p$ 是 $p$ 上所有余切矢量的集合，而流形 M 的所有点上的余切空间的笛卡尔积称为 M 的余切空间 $T^*M := \cup_{p\in M} T^*M_p$。同样，$2n$ 维的余切丛也具有自然的微分流形结构。陈省身指出通过在微分流形 M 上引入函数（芽），可以很自然地直接定义余切矢量和线性的余切空间（参见：[1] 第一章§2）。

微分流形上的标架丛是和切丛密切相关的。设 M 是 $n$ 维微分流形，所谓一个标架是指这样一个组合 $(p; \mathbf{e}_1, \cdots, \mathbf{e}_n)$，其中 $p$ 是 M 上一点，$\mathbf{e}_1, \cdots, \mathbf{e}_n$ 是流形 M 在点 $p$ 的 $n$ 个线性无关的切矢量。流形 M 上全体标架的集合记作 P。要在 P 中引进微分结构，使它成为光滑流形，并且使自然投影

$$\pi(p; \mathbf{e}_1, \cdots, \mathbf{e}_n) = p \quad (3)$$

是从 P 到 M 上的光滑映射，$(P, M, \pi)$ 称为 M 上的标架丛。标架丛 P 是与流形 M 的切丛 TM 相配的主丛，其纤维型和结构群都是一般线性变换群 $GL(n, \Re)$。

流动过程中流体内每一时空点都赋予一个宏观速度值，这使得流体流动的描述自然地具有矢量丛的构造。

## 2 微分形式、外积和外微分[1] [9]

对实数域上的矢量空间 V 定义实数值线性函数，这种全体实数值线性函数的集合是 V 的对偶空间 $V^*$，这种对偶关系是相互的。进一步利用张量积（多重线性函数）可以定义张量空间，称 $V_s^r = \underbrace{V \otimes \cdots \otimes V}_{r} \otimes \underbrace{V^* \otimes \cdots \otimes V^*}_{s}$ 为 $(r, s)$ 型张量，其中 $r$ 称为逆变阶数，$s$ 称为协变阶数，协逆变因子的排列也可以是交替出现的，$\dim V_s^r = n^{r+s}$。对协变（或逆变）张量进行对称化和反对称化分别得到对称张量和反对称张量，其集合记作 $S^r(V^*)$ 和 $\Lambda^r(V^*)$。

取矢量空间 V 为流形 M 在点 $p$ 的切空间 $TM_p$，于是在流形 M 的每一点 $p$ 有 $(r, s)$ 型张量空间 $T_s^r(p) = \underbrace{TM_p \otimes \cdots \otimes TM_p}_{r} \otimes \underbrace{T^*M_p \otimes \cdots \otimes T^*M_p}_{s}$。令 $T_s^r = \cup_{p\in M} T_s^r(p)$，在 $T_s^r$ 中引进拓扑、定义 $C^\infty$ 微分结构，就得到一个光滑流形 E，（陈讲义的原话是"我们要在 $T_s^r$ 中引进拓扑，使它成为有可数基的 Hausdorff 空间；进而可在 $T_s^r$ 上定义 $C^\infty$ 微分结构，使它成为一个光滑流形"）。对每一点 $p \in M$ 存在一个邻域 $U \subset M$，流形 E 局部同胚于积流形 $U \times T_s^r$，这样的 $T_s^r$ 称为流形 M 上的 $(r, s)$ 型张量丛。自然投影 $\pi: T_s^r \to M$ 把 $T_s^r(p)$ 中的元素映射到点 $p \in M$，且是光滑的满映射，称为丛投影，$T_s^r(p)$ 称为丛 $T_s^r$ 在点 $p$ 的纤维。

$n$ 维光滑流形 M 上的 $r$ 次外形式丛 $\Lambda^r(T^*M) = \cup_{p\in M} \Lambda^r(T^*M_p)$ 也是 M 上的张量丛，其光滑截面所成的空间记为 $A^r(M) = \Gamma(\Lambda^r(T^*M))$，$A^r(M)$ 中的元素称为 M 上的 $r$ 次外微分形式，它是光滑的反对称 $r$ 阶协变张量场，具有维数 $C_n^r$。所有阶次的外形式丛 $\Lambda(T^*M) = \cup_{p\in M} \Lambda(T^*M_p)$ 也是 M 上的矢量丛，$\Lambda(T^*M_p)$ 的维数为 $2^n$，$\Lambda(T^*M)$ 的截面空间 $A(M)$ 的元素称为 M 上的（外）微分形式，空间 $A(M)$ 的加法、数乘和外积成为一个外代数（exterior algebra，也称 Grassmann 代数，它是一种分次代数（graded algebra）：外积有一个重要性质，它使 $k$ 次外形式与 1 次外形式的积成为一个 $k+1$ 次外形式）。$r$ 次外微分形式 $\omega$ 限制在局部坐标为 $(u^1, \ldots, u^n)$ 的坐标域 U 上可表示为

$$\omega = \frac{1}{r!} a_{i_1 \ldots i_r} du^{i_1} \wedge \cdots \wedge du^{i_r}。 \quad (4)$$

其中 $a_{i_1 \ldots i_r}$ 是关于下指标反对称的 U 上的光滑函数。$\wedge$ 定义了外积运算 $\wedge: A^r(M) \times A^s(M) \to A^{r+s}(M)$，其中当 $r + s > n$ 时规定 $A^{r+s}(M)$ 为零。因为如下局部外微分运算 d 的存在（同时，微分形式是打算用作积分的被积式的，参见[8]苏竞存《流形的拓扑学》），使得微分形式在流形论中十分重要。

**定理 2.1** 设 M 是 $n$ 维光滑流形，则存在唯一的一个映射 d: $A(M) \to A(M)$ 使得 $dA^r(M) \subset A^{r+1}(M)$，并且满足下列条件：(1) 对任意的 $\omega_1, \omega_2 \subset A(M)$，$d(\omega_1 + \omega_2) = d\omega_1 + d\omega_2$；(2) 设 $\omega_1$ 是 $r$ 次微分形式，则 $d(\omega_1 \wedge \omega_2) = (d\omega_1) \wedge \omega_2 + (-1)^r \omega_1 \wedge d\omega_2$；(3) 若 $f$ 是 M 上的光滑函数，即 $f \in A^0(M)$，则 $df$ 恰是 $f$ 的微分；(4) 若 $f \in A^0(M)$，则 $d(df) = 0$。

**定义 2.2** $n$ 维光滑流形 M 称为可定向的，如果在 M 上存在一个连续的、处处不为零的 $n$ 次微分形式。如果在 M 上给定了这样一个微分形式 $\omega$，则称 M 是定向的；如果给出 M 的定向的两个微分形式彼此差一个处处为正的函数因子，则称它们规定了 M 的同一个定向。

下面进一步引入内积和李导数的定义以便应用。设 $\boldsymbol{V} = V^i \partial_i = V^i \frac{\partial}{\partial u^i}$ 是 $n$ 维光滑流形 M 上的矢量场，定义内积映射 $*$ : $TM \times \Lambda^r(M) \to \Lambda^{r-1}(M)$，使得

$$\boldsymbol{V} * \omega = \frac{1}{(r-1)!} V^j a_{ji_2 \cdots i_r} du^{i_2} \wedge \cdots \wedge du^{i_r}。 \quad (5)$$

利用内积和外微分运算，可以取 Cartan 的同伦公式定义微分形式的李导数如下：

$$\mathcal{L}_V \omega = \boldsymbol{V} * d\omega + d(\boldsymbol{V} * \omega) \quad (6)$$

满足 $d\alpha = 0$ 的微分形式称为闭微分形式 $\alpha \in \mathcal{C}(M)$；满足 $\alpha = d\beta$ 的微分形式称为恰微分形式 $\alpha \in \mathcal{E}(M)$。显然，$\mathcal{E}(M) \subset \mathcal{C}(M)$。彭加莱引理指出，当 M 是可光滑收缩到其中任一点时（比如星形域）有 $\mathcal{C}(M) \subset \mathcal{E}(M)$。还有一个可以改变微分形式次数的运算是霍奇星算子 $\star$ : $\Lambda^r(M) = \Lambda^{n-r}(M)$，使得

$$\star \omega = \frac{\sqrt{g}}{(n-r)!} e_{j_1 \cdots j_n} g^{j_1 i_1} \cdots g^{j_r i_r} a_{i_1 \cdots i_r} du^{j_{r+1}} \wedge \cdots \wedge du^{j_n}, \quad (7)$$

或

$$\omega \wedge \star \omega = g^{j_1 i_1} \cdots g^{j_r i_r} a_{j_1 \cdots j_r} a_{i_1 \cdots i_r} dv, \quad (8)$$

其中 $dv = \sqrt{g}\, du^1 \wedge \cdots \wedge du^n = \frac{\sqrt{g}}{n!} e_{j_1 \cdots j_n} du^{j_1} \wedge \cdots \wedge du^{j_n}$，$g_{ij} = \partial_i \cdot \partial_j$，$g_{ik} g^{kj} = \delta_i^j$，$g = det(g_{ij})$，$\delta_{j_1 \cdots j_r}^{k_1 \cdots k_r} = \frac{1}{(n-r)!} e^{k_1 \cdots k_r j_{r+1} \cdots j_n} e_{j_1 \cdots j_n}$，$e_{j_1 \cdots j_n} = \delta_{j_1 \cdots j_n}^{1 \cdots n} = \begin{vmatrix} \delta_{j_1}^1 & \cdots & \delta_{j_n}^1 \\ \vdots & \ddots & \vdots \\ \delta_{j_1}^n & \cdots & \delta_{j_n}^n \end{vmatrix}$。定义霍奇星算子时需要知道流形的定向，它可以用一组有序的协变基表示，比如 $\left\{\frac{\partial}{\partial u^1}, \cdots, \frac{\partial}{\partial x^n}\right\}$，基的轮换或偶数次置换不改变定向，定向用微分形式表示更为简洁，它就是体元形式 $du^1 \wedge \cdots \wedge du^n$。二维和三维右手系取向的图像表示如图 2-1。

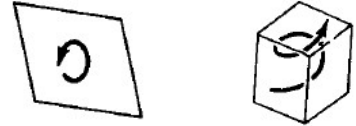

图 2-1 二维、三维定向的图形表示
**Figure 2-1** Graphical representation of 2D and 3D orientations

## 3 微分形式的几何表征[10]

速度场矢量表示的积分图像是流线，速度越大流线越密（如图 3-1 所示）；这一图像对应的微分形式是流量通量形式 **F**，这个扭 2-形式表示垂直通过微面的有向线的密度（如图 3-2 所示）。速度场的微分 1-形式表达可以看作是对流量通量形式进行霍奇星运算得到的：$\widetilde{\mathbf{V}} = \star \mathbf{F}$。它给出处处与速度矢量垂直的微面，相邻微面之间距离的倒数反映了速度的大小（"即定义面的场的梯度"），与流线的整体性不同，这些处处垂直于速度矢量的面通常不能形成一个整体（"不可积"：只有速度场有势或速度 1-形式是全微分的情况才可以）。速度 1-形式的几何图像如图 3-3 所示，它的大小反映单位长流线内微面的密度，箭头"<"表示方向。

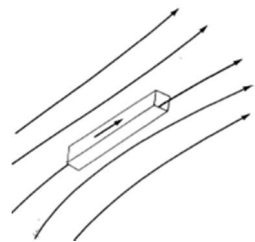

图 3-1 速度矢量的图像—流线
**Figure 3-1** Image of velocity vector — streamline

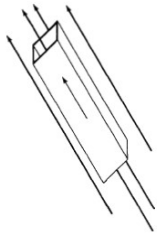

**图 3-2** 流量通量形式的图像
**Figure 3-2** Image of flow flux 2-form

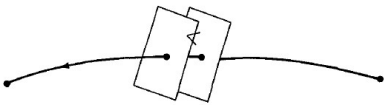

**图 3-3** 速度 1-形式的图像
**Figure 3-3** Image of velocity 1-form

有时微分 1-形式即使是全微分也可能是不平凡的，比如极坐标中角度变量的微分只有挖去原点才是处处有意义的，它的几何图像如图 3-4 所示。点涡诱导的无旋速度场是另一个实例。在去掉原点之后，每个闭微分形式都是恰微分形式这一论断就不再成立了，这涉及到流形的整体拓扑（上同调）。于是，微分形式具有一种奇妙的特性，一方面它是局部的，另一方面它的解析性质与内禀拓扑息息相关！闭微分形式空间与恰微分形式空间的商空间称为 De Rham 上同调群，上同调群的维数称为贝蒂（Betti）数[11]。如果所有闭微分形式都能写成恰微分形式，则贝蒂数等于零；反之亦然。上述通过微分形式分析流形拓扑的思路有个对偶，即也可以通过微分形式在边界的积分（闭路积分）来分析流形的拓扑，涉及单形、单纯复形和（下）同调等概念，兹不详述。

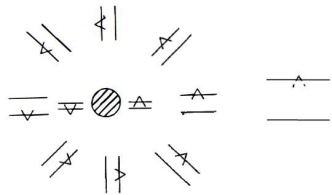

**图 3-4** dθ的图像
**Figure 3-4** Image of dθ.

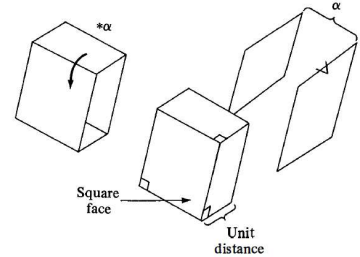

**图 3-5** 星运算把 1-形式α变成 2-形式
**Figure 3-5** Star operation changing 1-form into 2-form

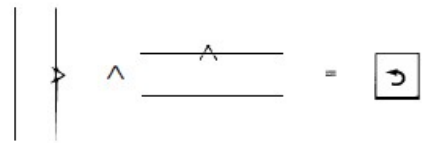

**图 3-6** 二维 1-形式外积得到 2-形式
**Figure 3-6** Exterior product obtaining 2-form from 1-form in 2D space

在已知定向情况下的微分形式称为扭微分形式，记作$(\omega, \Omega)$，霍奇星算子也有与取向结合的考虑，称为扭霍奇星算子，记作$(\star, \Omega)$。$(\star, \Omega)\omega = (\star \omega, \Omega)$，$(\star, \Omega)(\omega, \Omega) = \star \omega$。微分形式和扭微分形式的图像表示有所不同，如图 3-7—3-9 表示。

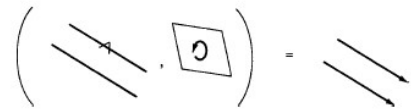

**图 3-7** 二维扭 1-形式
**Figure 3-7** Twisted 1-form in 2D space

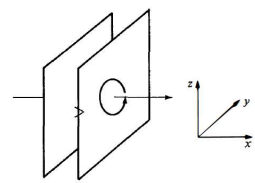

**图 3-8** 三维 1-形式和扭 1-形式的对应
**Figure 3-8** Correspondence between 1-form and twisted 1-form

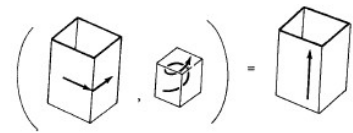

**图 3-9** 三维扭 2-形式

**Figure 3-9** Twisted 2-form in 3D space

速度 1-形式沿闭合回路的积分称为环量，这是理想流体力学的一个重要概念。速度 1-形式的外微分得到涡量 2-形式，速度 1-形式与涡量 2-形式的外积得到螺旋度 3-形式即三维体形式。体形式的图像与取向的图像相同（见图 2-1）：二维是一个涡、三维是一个螺旋涡。因此，三维涡量 2-形式的图像有两种，考虑取向时是图 3-2，不考虑取向时见图 3-5。

## 4 经典流动理论的微分形式表述[12][10]

记处处一致的笛卡尔坐标系的自然基为 $\left\{\frac{\partial}{\partial x^1}, \frac{\partial}{\partial x^2}, \frac{\partial}{\partial x^3}\right\}$ 或 $\{\mathbf{e}_1, \mathbf{e}_2, \mathbf{e}_3\}$，微分形式基为 $\{dx^1, dx^2, dx^3\}$，有体元形式 $dv = dx^1 \wedge dx^2 \wedge dx^3$ 和面元形式 $da_i = \frac{1}{2}e_{ijk}dx^j \wedge dx^k, i=1,2,3$。引入质量形式 $m = \rho dv$、动量形式 $\mathbf{w} = \rho\mathbf{V}dv$、压力形式 $\mathbf{p} = p\mathbf{e}_i\delta^{ij}da_j \wedge dt$ 和粘性力形式 $\boldsymbol{\sigma} = \sigma^{ij}\mathbf{e}_i da_j \wedge dt = \mu \star d\mathbf{V}$（此处 μ 是指粘性系数，也即，我们取最简单各向同性线性本构关系；采用霍奇星算子是因粘性力与速度空间微分是一对广义力和广义流），经典流动理论（连续方程和动量方程：第七节将介绍邹文楠的独特流动场论）可以用微分形式表示为[1]

$$(\partial_t + \mathcal{L}_V)m = 0, \quad (9)$$

$$(\partial_t + \mathcal{L}_V)\mathbf{w} = d(-\mathbf{p} + \mu \star d\mathbf{V})。 \quad (10)$$

利用（7）式，动量方程（8）有以下常规写法：

$$\rho(\partial_t + V^j\partial_j)V^i = -\delta^{ij}\partial_j p + \mu\nabla^2 V^i. \quad (11)$$

若引入时空 3-形式

$$\omega = \rho(dx^1 - V^1 dt) \wedge (dx^2 - V^2 dt) \wedge (dx^3 - V^3 dt), \quad (12)$$

经典流动理论又可表述为

$$(\partial_t + \mathcal{L}_V)m = 0 \Leftrightarrow d\omega = 0, \quad (13)$$

$$(\partial_t + \mathcal{L}_V)\mathbf{w} = d(-\mathbf{p} + \mu \star d\mathbf{V})$$
$$\Leftrightarrow d(\mathbf{V}\omega + \mathbf{p} - \mu \star d\mathbf{V}) = 0. \quad (14)$$

速度是经典流动理论的核心概念，用它可以定义各种有意义的微分形式，除了独立地作为矢量场、与质量形式结合构成动量形式外，还有流量通量形式 $F = V^i da_i$ 及与之对偶的速度 1-形式 $\tilde{V} = \delta_{ij}V^i dx^j$。基于速度 1-形式，定义它的升阶序列

$$I_1 = \tilde{V}, I_2 = \tilde{\Omega} = d\tilde{V} = \Omega^i da_i, I_3 = \tilde{H} = \tilde{V} \wedge \tilde{\Omega} = Hdv。 \quad (15)$$

显然，按照构造方法 $I_{2k+1} = I_1 \wedge I_{2k}, I_{2k+2} = dI_{2k+1} = I_2 \wedge I_{2k}$，三维空间的速度 1-形式在任一点上满足 $I_{K(P)} \neq 0, I_{K(P)+1} = 0$ 的 Darboux 级 $K(P)$ 不大于 3。微分 1-形式 $\tilde{V}$ 在域 $\wp$ 上的 Darboux 级定义为 $K(\tilde{V}, \wp) = \max_{P \in \wp} K(P)$。设 $\phi, \psi, \varphi$ 是上互相独立的函数（两个标量函数互相独立 $\Leftrightarrow df \wedge dg \neq 0$），且 $\psi$ 恒为正值，根据 Darboux 定理，如果 $I_3 \neq 0$，有表示 $\tilde{V} = d\phi + \psi d\varphi$，如果 $I_2 \neq 0, I_3 = 0$，有 $\tilde{V} = \psi d\phi$，如果 $I_2 = 0$，有 $\tilde{V} = d\phi$。即速度场分别需要 3 个、2 个和 1 个独立函数表示。另外，有以下积分关系 $\Gamma = \oint \tilde{V} = \iint \tilde{\Omega}, \oiint \tilde{\Omega} = 0$。

利用速度 1-形式，根据协变导数和 Lie 导数间的关系，还可以改写（8）式为

$$(\partial_t + \mathcal{L}_V)\tilde{V} = -\frac{dp}{\rho} + d\left(\frac{V^2}{2}\right) + \nu\nabla^2\tilde{V}. \quad (16)$$

由于外微分和李导数可以交换，可得不可压流动涡量 2-形式的方程为

$$(\partial_t + \mathcal{L}_V)\tilde{\Omega} = \nu\nabla^2\tilde{\Omega}. \quad (17)$$

因为涡量 2-形式是恰微分，不考虑粘性时，上述方程意味着涡量 2-形式是柯西不变量。

## 5 外微分描述 Euler 方程经典解基本性质和爆破问题一例简介（详情咨询：jz@sccfis.org）

本节作为第一个例子，我们简要讨论一个最近用于研究 Euler 方程爆破问题的修正 Euler 模型[13]，作为一例来展示外微分在揭示动力学基本性质，尤其是手性相关方面（如，局部螺旋度 – local

---

[1] 对于下面式子中右手边为零的情况，文献一般称为"李对流（Lie-advection）"或"李输运（Lie-transport）"。这是标量问题术语的沿袭。由于流动中 "对流"和 "输运"的含义就不尽相同，不同场合下它们本身所指也会略有不同，容易引起混乱。因此，我们尝试建议用一个既贴切又全新的术语"李携带（Lie-carry）"，但有时也沿用通常惯例。

helicity/spirality[14]）的便利应用，以及对湍流研究新思维的启发。一般读者可以粗略浏览以获得一些本文主旨的感觉，并不影响后续章节的阅读；有兴趣的读者则可在此基础上继续仔细研究相关文献。

传统（理想）流体力学讨论的动力学不变量主要是两种。一个是局部的，或者说几何的[2]，随流动不变的，也叫物质不变性，或现代微分几何对一般的张量引入的李（Lie）不变：熟知的例子有被动标量输运和涡冻结（及其派生出的开尔文-亥姆霍兹诸定理），与手性相关的是局部螺旋度；另一个是全局的，对流域积分的一个量（所以也如哈密顿力学中叫运动积分/常数、初积分等）。据我们所知，在这两个方面，目前的非线性数学知识并不能告诉我们关于 Euler 方程的全部相关信息：我们不知道除了定义动力学的质量、动量和能量守恒到底还有其它多少独立（不由它们线性组合）的守恒量。反过来，从已知的守恒律我们能获得流动的什么物理或力学性质，也是不断地在被发现。[3]这方面的任何一个进展都是根本性的，对于揭示湍流的基本性质也极具威力。

我们知道三维 Euler 方程解的全局存在唯一性问题至今还是悬（赏）而未决的问题，而二维全局解存在唯一性的证明则充分利用了不变量：二维涡量的任意函数的积分都是常数，充分刻画了每一余伴随轨道，但也仅限于此，并不代表系统可积[15]。对于三维问题，我们只知道两个初积分，动能和螺度，而它们似乎并不足以保证全局解的存在：Tao[13]最近就构造了修正的 Euler 方程，形式地同样保持能量和螺度，以及开尔文环量守恒定理。但是，其初始光滑数据在有限时间内爆破。下面我们结合其它要讨论的内容稍微展开介绍。

经典不可压缩流理论中构型变化由特殊微分同胚（SDiff）群 $G$ 描述，其单位元 $e$ 处切空间 Lie 代数 $g=T_eG$ 就是速度，每一速度矢量空间有一个对偶的外微分形式矢量空间$g^*$，它们之间由度规g联系。考虑将（17）式无粘情形中的$\mathcal{L}_V$换成$\mathcal{L}_U$

$$(\partial_t + \mathcal{L}_U)\tilde{\Omega} = 0, \quad (18)$$

并选择U和V的关系为

$$\tilde{U} = d^\star A\tilde{\Omega}. \quad (19)$$

如前面，$d\tilde{V} = \tilde{\Omega}$。文献[13]可能是为了简化符号，将矢量和外微分形式在表达式中直接转换，称与这里的 $A$ 相应的复合算子为矢量势算子。可以令$A$满足适当条件，使得修正的系统具备前面提到的 Euler 方程的常见性质。这是"曲线救国"的思路，也是 Kraichnan 的湍流解析模型（从 DIA 到后面的各种演变及简化版本）的思想出发点。Tao 构造了三个具体模型，证明它们的初始光滑解都在有限时间内爆破，由此推论，要想证明存在全局解，应该还需要 Euler 方程的其它信息。显然构造模型模拟湍流也是这个道理，需要尽量保持原方程的结构性信息。这是文献[14]涉及的一个思想主题：一方面，如果要在流动中控制保持某种几何结构，尤其是高阶[4]如局部螺旋度（local helicity/spirality）这样的几何对象，如何实现（工程问题中，我们确实常常有这样的问题，比如，局部的旋涡分离流，它们对于噪声和壁材料本身相关的力、热和声等问题意义重大）；另一方面，对于非理想因素（如，粘性耗散）对局部几何结构冻结于流体的破坏，如果还是要从"冻结"【于其它（虚拟的）流动[5]】角度来看，又如何？

虚拟路径在经典意义上的确定性可能有更深刻的意义和应用价值[18]：两个可能相关的例子是，量子力学的 Feynmann 路径积分（或者说随机过程中的 Chapman-Kolmogorov 方程）和 Iyer 和 Constantin[19]最近将分子粘性用布朗运动刻画的随机冻结描述：对

---

[2] 除了前面介绍的外微分形式的几何图像含义，另一个原因是，前面介绍的 Lie 导数可视为由坐标变换而来，李不变意味着只涉及坐标变换。
[3] 全局守恒量方面，比如，李政道就根据能量守恒和遍历性假设，通过统计计算为 A. N. Kolmogorov 作为其"公理化"的小尺度湍流各向同性假设提供依据；R. H. Kraichnan 论证弱可压缩湍流中噪声对湍流衰减的延缓作用，进一步对三维不可压缩能量和螺度的正向级串方向做出解释，尤其是利用二维不可压缩流动中能量和拟涡能同时守恒的性质，成功预测了能量可以反向级串到大尺度；朱建州指出螺度（关于此物理量的讨论见下一节）可用于降低湍流噪声（密度摸和压缩摸等），等等：见最后一节讨论。
[4] 前面方程（17）给出的只是到 2-形式，高阶的或其它低阶的需要另外构造。
[5] 就像现代偏微分方程理论中的广义解，我们也可能将虚拟速度视为"广义的"（具体含义可根据不同情况而定，比如，可以是不连续、非确定性或随机的，等等非经典意义上的）路径上的冻结速度。

于"经典"路径，前者（量子力学中）是所有"虚拟"路径中拉格朗日作用量取极值的那条，后者（Navier-Stokes 规则流动）是"虚拟"路径的平均。这是非平凡和"有用"的，因为引入虚拟速度的"实用"意义是为一般（非理想）流动 --- 尤其是湍流 --- 的描述和模拟另辟（粒子观点的）蹊径。

对于规则流动，即便在中性流体中，文章[17]说明在两维两分量与一维三分量耦合的流动中也可能分解出涡量，它冻结于另一个（子）流动。最近杨越课题组[16]就携带涡矢量的经典虚拟速度在非理想流体和磁流体中展示了存在唯一性问题。对于一般的几何对象，Lie 携带的虚拟速度的存在唯一性问题没有一个统一的处理方法。文稿[18]的动机之一，就是有感于现有的流动全局拓扑描述方法在一般流动中面临的困难。比如文章[14]强调，传统采用 Clebsch 势函数的 Pfaff 描述就有全局存在/可积性的问题，而相关的局部螺旋度刻画了其不可积性的程度，所以要控制局部螺度。正如文章[16]所展示，经典虚拟速度和涡面场一样，一般情况下也会碰到困难（因而在计算机模拟中需要数值的规则化处理），但提供了新的视角，尤其是有意义明确的输运图像。

（18,19）式引入$\tilde{U}$作为虚拟的冻结速度（1-形式），事实上构建了一个双流体模型。虽然（19）式已经显式地给出了虚拟速度，但是为更好地理解和比较动力学，尤其是局部几何结构，"刻意地"推出动力学方程（以显示其局部微分结构）也是有意义的。为此，我们对（19）式两边对时间求导。时间导数可以有很多种，可以是空间坐标固定的 Euler 局地导数，也可以是随（任意某个流）体的 Lagrange 导数。作为示意，文章[14]中假设当地度规不随时间变化 --- 欧几里得度规显然不随时空变化 --- 对（19）式两边取局地导数，它与空间算子可交换；假设$A$也是简单的可与其交换的空间作用算子，带入（18）式就有

$$\partial_t \tilde{U} + d^\star A \mathcal{L}_U d\tilde{V} = 0 \quad (20)$$

由于左边第二项中李导数算子与其他算子一般来说不可以交换，所以这个虚拟流动的 Lie 代数结构看来并不明朗（主要在于余微分算子与矢量势算子的直观含义不明显）。（20, 16）两式显示这两个流动结构非常不对称。文献[14]还讨论了等离子体双流体模型，后者具有非常对称的结构，也是描述等离子体流动最全面的流体模型，由它可通过各种简化而得到通常所谓的（单流体）磁流体，我们这里不展开讨论。由于$g^*$与商空间$\Lambda^1/d\Lambda^0$同构，也就是说其元素可视为由所有形如$\tilde{V} + df$（$f$为任意某函数）的 1-形式构成的陪集（coset）$[\tilde{V}]$。所以对应于（16）式的广义 Euler 方程为

$$(\partial_t + \mathcal{L}_V)[\tilde{V}] = -df. \quad (21)$$

为避免对通常流体力学读者造成复杂难懂的印象，此处保持最简，暂不讨论 Lie-Poisson 结构和 Hamilton 方程，以及余伴随轨道（coadjoint orbit）问题。最后一节会回来简要点到。

换句话说，我们实际上可取特殊的$\tilde{W}\epsilon[\tilde{V}]$，使得

$$(\partial_t + \mathcal{L}_V)\tilde{W} = 0. \quad (22)$$

这样$\tilde{W} \wedge \tilde{\Omega}$就也是李不变的了，它就是 Oseledets[20] 构造的局部螺旋度（spirality）[6]。我们也可以对（19）式两边取随体导数，从而引入李输运结构，但是通常也会留下一些其它残余项，模糊了几何意义。如前面提及，如何找到某个（虚拟的）流，使得目标几何结构对象随其的导数具有清晰的李输运结构，这是另一个问题。当然，（22）和（16）之间的"中间道路"是同时调整选择合适的虚拟流动和几何对象，使得既物理图像清晰、"有用"又数学描述干净优美。这可能是解决复杂一点的流动问题的思路，有待进一步在具体问题中去尝试。

---

[6] 如此构造高阶几何不变量的技术被 Besse 和 Frisch[14]反复用在不同例子中，以展示他们的广义柯西不变量方程结果。文章[17]中所谓"略微的"推广之一指的是，要形成高阶几何不变量并不需要这么强的条件。比如，（22）式右边如果不为零的 S，只要 $S \wedge \tilde{\Omega} = 0$，那$\tilde{W} \wedge \tilde{\Omega}$还是一个几何不变量。其实我们现在发现，文献[4]提到的 Ertel 势涡（potential vorticity）对于非正压气体的绝热流动过程就是一个例子；也即，我们（16）式中的$d(\frac{dp}{\rho})$不为零并不影响 Ertel 势涡（3-形式及其 Hodge 对偶的标量）作为几何不变量。文章[14]中"更基本的"推广是指，如果这个 S 以及像（16）式那样其它配套的方程中右边不是像$\frac{dp}{\rho}$一样是流动内禀的，而是视为外来（如边界）的驱动或耗散，我们也可以适当地构造几何不变量以应用 Besse 和 Frisch[4]的广义柯西不变量方程，并可能作为湍流（拉格朗日视角下的）理论的约束。

为从（18）式出发构造三阶局部螺旋结构，我们需要首先得到速度一阶外微分形式的方程。由于外微分和第一个时间导数及第二个李导数可交换，（18）可写为$d(\partial_t + \mathcal{L}_U)\tilde{V} = 0$，假设没有上同调（比如，区域单连通），Hodge 分解即给出

$$(\partial_t + \mathcal{L}_U)\tilde{V} = d\tilde{P}. \quad (23)$$

上面$d\tilde{P}$与原始"真实"动力学相应项并无直接联系。剩下的工作和前面得到（22）式下面给出的李不变量$\tilde{W} \wedge \tilde{\Omega}$是一样的。继而还可以写出对应于虚拟流动 Lagrange 轨迹的广义 Cauchy 不变量方程[14]。这个结果的意义在于具体给出构造"虚拟"李不变高阶几何对象的例子。

关于更细节的性质讨论和解的爆破证明[13]，我们这里不展开讨论。文章[14]讨论各种局部几何结构以及作为湍流模型的约束问题，所以也给出了虚拟流动三阶几何不变量（局部螺旋度）所对应的柯西不变量方程。但是，包括对可压缩流和等离子体流[7]的讨论，这里只能点到为止。对于湍流精准刻画理论问题，我们最后强调的是，各阶几何结构（相应地可以有对涡线的推广[21]）约束的必要性和数学上外微分表述的控制手段。为更加明确这一点，作为参考实例，读者可以考虑"广义拉格朗日平均"理论[3]。

# 6 流体力学中的螺度(helicity)以及整体拓扑性质

## 6.1 Helicity 的引进以及流体纽结/链环的（自）缠绕数之间的关系

螺度的思想最早来源于十九世纪的 Helmholtz 和 Kelvin[32]（下一小节我们会对此再深入解释说明），此后 1958-1969 年在磁流体力学和理想流体动力学中分别由 Woltjer[51]和 Moffatt[38]重新加以发扬光大。Woltjer 给出了 Helicity 的三维积分定义（参下文），Betchov 最早[25]讨论（中性流体）湍流中螺度效应问题。Moffatt[38]给出了它与链环分量的高斯互缠绕数（Gauss linking number）之间的关系。之后，1992 年 Moffatt 和 Ricca[39][46]又进一步给出了它与纽结族的自缠绕数、互缠绕数之间的关系。以下回顾内容可参看文献[22]。

考虑一个充满了理想流体的三维单连通流形，为简单起见我们考虑欧氏空间$\mathbf{R}^3$ 中的一个区域$V$。涡旋动力学中的运动学螺度（kinetic helicity）以流体速度场为基本场来定义：

$$H = \int_V \mathbf{u} \cdot \boldsymbol{\omega}\, d^3x, \quad (24)$$

其中$\mathbf{u}$是流体速度场，$\boldsymbol{\omega}$是涡旋度（vorticity），$\boldsymbol{\omega} = \nabla \times \mathbf{u} = curl\, \mathbf{u}$，这里$curl$是旋度算子。为简单起见，要求$\mathbf{u}$在$V$内处处满足$\nabla \cdot \mathbf{u} = 0$，而$\boldsymbol{\omega}$在边界$\partial V$上满足$\boldsymbol{\omega} \cdot \hat{\mathbf{n}} = 0$，其中$\hat{\mathbf{n}}$是$\partial V$的法向。$\boldsymbol{\omega}$满足$\nabla \cdot \boldsymbol{\omega} = 0$。

磁流体力学中的螺度（magnetic helicity）也有类似的定义。即代换$\mathbf{u} \to \mathbf{A}$，$\boldsymbol{\omega} \to \mathbf{B}$，其中$\mathbf{A}$是磁矢势，$\boldsymbol{\omega}$是磁场强度。那么就有

$$H = \int_V \mathbf{A} \cdot \mathbf{B}\, d^3x, \quad (25)$$

用外微分语言来重新表述上面的定义知：速度场$\mathbf{u}$是一个 1-形式，$\mathbf{u} = u_i dx^i$；$\boldsymbol{\omega}$，作为一个 1-形式，本身定义为一个 2-形式的 Hodge 对偶，$\boldsymbol{\omega} = *(d\mathbf{u})$。从而(24)式就可以表达为一个外积（wedge product）：

$$H = \int_V u \wedge du. \quad (26)$$

如果反过来以$\boldsymbol{\omega}$为基本场，亦可把上述定义写成

$$H = \int_V \boldsymbol{\omega} \cdot curl^{-1} \boldsymbol{\omega}\, d^3x, \quad (27)$$

其中$\mathbf{u} = curl^{-1} \boldsymbol{\omega}$由 Biot-Savart 定理来实现。

$$\mathbf{u}(\vec{x}) = \int_{V'=V\setminus\{\vec{x}\}} \frac{\boldsymbol{\omega}(\vec{x}') \times (\vec{x}-\vec{x}')}{|\vec{x}-\vec{x}'|^3}\, d^3x'. \quad (28)$$

文献[22]P125 定理 1.15 指出：$H$的取值并不依赖于$u$取何种形式，而实质上取决于$\boldsymbol{\omega}$。（此定理的证明从略；读者可参看文献[22]P126。）这一点与物理中的通常认知是一致的，即$\boldsymbol{\omega}$作为场强来说是可观测量，而$u$从场论的角度看是一种矢量势。定理的一个直接结果是如下的重要推论，即 Helicity Invariance Theorem：设流形$M$上有一个无源矢量场$\boldsymbol{\omega}$。则由$\boldsymbol{\omega}$所给出的$H$，在$M$的任意保体积微分同胚（diffeomorphism）变换之下是守恒的。具体来说就是，对一个单连通带边流形

---

[7] 可以通过引入半直积来将不可压缩经典流动的李群李代数描述拓展到这两种情形[15]。

$M$来说，若一个无散（divergence-free）矢量场和边界$\partial M$相切，那么该场的螺度$H$在所有针对$M$的保体积、保持上述相切条件不变的微分同胚变换之下是一个守恒量。[8]从拓扑学的角度，文献[22]也将$H$称为Hopf不变量，即Hopf映射度。

下面讨论$H$作为拓扑不变量与流体纽结（knot）/链环（link）族的（自）缠绕数之间的关系。Moffatt 在 1969 年指出[38]，当不考虑纽结的自身缠绕时，$H$可以表达为流体中的纽结/链环族的高斯互缠绕数之总和。

考虑理想磁流体条件之下的一个磁链环$L$，当中包含有两个或更多个相互缠绕的磁通管。设第$i$根磁通管画在平面上后，其中心线为$C_i$，所携带的磁通为$\Phi_i$。如下图。

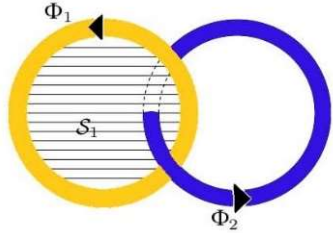

**图 6-1** 一个简单的磁链环，包含两个磁通管。
**Figure 6-1** A simple magnetic link, containing two magnetic tubes.

那么这个磁流体系统的 Helicity 就可用链环的拓扑数，高斯互缠绕数，来表达：

$$H = \sum_{i \neq j} \Phi_i \Phi_j Lk(C_i, C_j), \quad (29)$$

其中$Lk(C_i, C_j)$是$C_i$和$C_j$之间的高斯互缠绕数。

证明简述如下。由上文 Biot-Savart 定理知

$$A(\vec{x}) = \int_{V'=V\setminus\{\vec{x}\}} \frac{B(\vec{x}')\times(\vec{x}-\vec{x}')}{|\vec{x}-\vec{x}'|^3} d^3x'. \quad (30)$$

将之代入磁螺度的定义，就有

$$H = \int_V A(\vec{x}) \cdot B(\vec{x}) d^3x = \int_V \left(\int_{V'=V\setminus\{\vec{x}\}} \frac{B(\vec{x}')\times(\vec{x}-\vec{x}')}{|\vec{x}-\vec{x}'|^3} d^3x'\right) \cdot B(\vec{x}) d^3x, \quad (31)$$

也即

$$H = \int_V \int_{V'=V\setminus\{\vec{x}\}} \frac{B(\vec{x})\times B(\vec{x}')}{|\vec{x}-\vec{x}'|^3} \cdot (\vec{x}-\vec{x}') d^3x' d^3x. \quad (32)$$

与前文(24)式类似地，取$B$场在边界$\partial V$处满足$B\cdot\hat{n}=0$，就可把$B(\vec{x})$和$B(\vec{x}')$场表为

$$B(\vec{x}) = \Phi_i \hat{n}(\vec{x})\delta^3(\vec{x}-\vec{x}_i),$$
$$B(\vec{x}') = \Phi_j \hat{n}(\vec{x}')\delta^3(\vec{x}'-\vec{x}_j), \quad (33)$$

其中$\vec{x}_i$和$\vec{x}_j$分别是磁通管$C_i$和$C_j$上的坐标，$\Phi_i$和$\Phi_j$分别是$C_i$和$C_j$的磁通量。由于$\delta$-函数的缘故，$\hat{n}(\vec{x})$和$\hat{n}(\vec{x}')$就分别成为$C_i$和$C_j$上的单位切矢量。那么，将$B$场的表达式代入上式，立刻得到

$$H = \sum_{i\neq j} \Phi_i \Phi_j \oint_{C_i} \oint_{C_j} d\boldsymbol{l}_i \times d\boldsymbol{l}_j \cdot \frac{(\vec{x}_i-\vec{x}_j)}{|\vec{x}_i-\vec{x}_j|^3}. \quad (34)$$

注意到高斯互缠绕数$Lk(C_i, C_j)$恰为上述积分，即得证

$$H = \sum_{i\neq j} \Phi_i \Phi_j Lk(C_i, C_j),$$

其中 $Lk(C_i, C_j) = \oint_{C_i} \oint_{C_j} d\boldsymbol{l}_i \times d\boldsymbol{l}_j \cdot \frac{(\vec{x}_i-\vec{x}_j)}{|\vec{x}_i-\vec{x}_j|^3}$ 。 (35)

高斯互缠绕数的计算是简单的。它不仅可以用上面的线积分来得到，作为拓扑数还可用如下的简单代数计数来求取：

$$Lk(C_i, C_j) = \frac{1}{2}\sum_{r\in\{C_i\sqcap C_j\}} \varepsilon_r, \quad (36)$$

其中$\{C_i\sqcap C_j\}$指$C_i$和$C_j$的交点集合，$r$指当中的一个。$\varepsilon_r$指交点$r$的代数指标，定义如下：将流体纽结/链环画到平面上，也即选定一种方式把三维曲线投射到二维上去，于是每个交叉点的情形就固定下来（—— 这是在二维当中反映三维景象的一种可行方案）。那么，每个交叉点有两种情况，如图6-2：

---

[8] 这可以推出螺度守恒，但是不是螺度守恒的必要条件。事实上，甚至在非理想系统中螺度可以是守恒的，而其拓扑可以变化。我们将在最后一节再回到这一点。另外，前面有讨论到，一个流动中可能有不止一个螺度守恒。

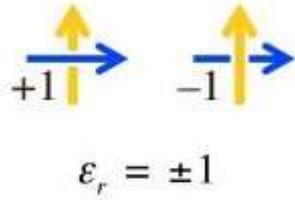

**图 6-2** 对交叉数的代数定义：Over-pass，即从上线转到下线的最小转角为逆时针转动，记为$\varepsilon_r = +1$；Under-pass，即最小转角为顺时针转动，记为$\varepsilon_r = -1$。

**Figure 6-2** This Algebraic definition for the crossing number: Over-pass, i.e., the case that the minimal angle of rotation from the upper line segment to the lower line segment is anti-clockwise. It is denoted by $\varepsilon_r = +1$; Under-pass, i.e., the case that the minimal angle of rotation is clockwise. It is denoted by $\varepsilon_r = -1$.

这样一来之前 Helicity 很复杂的三重积分计算，就可以用简单的代数计数来代替。举例如下：

☐ 示例一，上面的图 6-1。容易看出两个交叉点的代数指标都是+1。那么根据公式有$Lk(C_i, C_j) = \frac{1}{2}(+1 + 1) = 1$，恰为两个磁通管的缠绕数。

☐ 示例二，如图 6-3。其中两个交点的代数指标，一为+1，一为−1，故总和为零，

$$Lk(C_i, C_j) = \frac{1}{2}(+1 - 1) = 0。$$

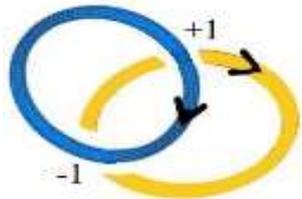

**图 6-3** 两个平庸圈，互不缠绕。考察其交点的代数指标，一为+1，一为−1，互相抵消。

**Figure 6-3** Two trivial circles without tangling. The algebraic indices of their intersection points are +1 and −1 separately, hence the indices cancel each other.

**图 6-4** 一组三个平庸环（左），Borromean rings（中）和 Whitehead links（右）。平庸环显然互缠绕为零；后二者，无论怎样规定各分量（用不同颜色标定）的定向，其总高斯互缠绕数都等于零。

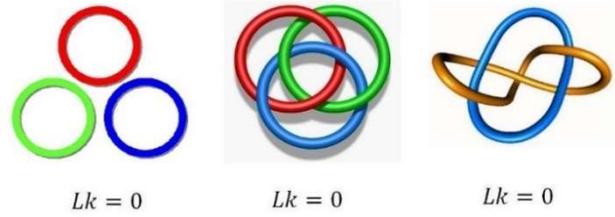

**Figure 6-4** This figure A group of three trivial circles (*left*), Borromean rings (*middle*) and Whitehead links (*right*). The mutual linking number of the trivial circles are obvious zero; for the latter two items, their mutual linking numbers both vanish, no matter how to choose the orientations of the components in the links.

☐ 示例三，一组三个平庸环、Borromean rings 和 Whitehead links，如图 6-4。它们各自的总高斯互缠绕数都为零。

1992 年 Moffatt 和 Ricca 又进一步讨论了纽结的自缠绕情况[39][46]。可以证明 Călugăreanu-White 定理：考虑一个闭合带$R = R_\epsilon(C, C^*)$。这里$C$是带子的一条边缘，可视为带子的中心线，描述的是带子的整体弯曲性状。$C^*$是另一条边缘，可视为$C$的 framing，简单理解为阴影，Framing 宽度是无穷小$\epsilon$。$C^*$与$C$搭配起来描述带子自身的拧转（twisting）。那么可以证明 Călugăreanu-White 公式，自缠绕数（self-linking number）

$$Sl(R) = Wr(C) + Tw(R), \quad (37)$$

其中$Wr(C)$是环绕数或绕数（writhing number），由整体弯曲$C$决定；$Tw(R)$是拧转数或拧数（twisting number），由带子自身的拧转情况决定。定理的证明从略，当中要讨论拐点处的极限情况。而$Tw(R)$又有内部分解，

$$Tw(R) = T(C) + N(R), \quad (38)$$

其中$T(C)$称为总挠率（total torsion），而$N(R)$是内禀拧转（intrinsic twisting）。这样一来，可以证明 Helicity 与自缠绕数、缠绕数的关系：

$$H = \sum_i \Phi_i^2 Sl(C_i) + \sum_{i \neq j} \Phi_i \Phi_j Lk(C_i, C_j). \quad (39)$$

证明从略，参看[39][46]。

自缠绕数的计算也可以采取代数计数的简单算法。例如图 6-5 所示的左手和右手三叶结（trefoil knot），它们的自缠绕数分别是$-\frac{3}{2}$和$+\frac{3}{2}$。

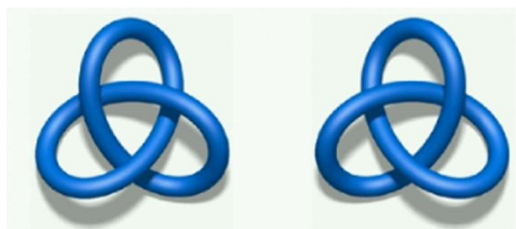

**图 6-5** 左手三叶结（左）和右手三叶结（右）。左手结的自缠绕数为$-\frac{3}{2}$，右手为$+\frac{3}{2}$。
**Figure 6-5** Left-handed trefoil knot (left) and right-handed trefoil knot (right). The self-linking number of the left-handed one is $-\frac{3}{2}$, while that of the right-handed is $+\frac{3}{2}$.

## 6.2 拓扑学基本概念·拓扑示性类·Helicity 与陈-Simons 形式的对应

6.1 节开头所言"螺度的思想最早来源于十九世纪的 Helmholtz 和 Kelvin[32]"对于螺度为理想流动所保持应如下理解：Helmholtz-Kelvin 定理说涡量是冻结于流体的，说速度沿任一（闭）物质线的环量不变。这样，如果我们取物质线就是涡线，那么不变的环量就是螺度。螺度动力学守恒其实有对应严格而深刻的数学理论，那就是 Noether 第二定理。Euler 方程对 Lagrange 流体微团有重标记对称性，这个对称性对应有一个不变积分，取重标记变换沿着某个环，这个不变积分就是 Kelvin 环量，而取这个环为涡环，这个不变积分就是螺度。螺度是一个拓扑不变量（文献[22]称 Hopf 不变量），下文还将把它与陈-Simons 次级示性类相比较，因此有必要回顾一些拓扑学的基本概念（参看[26][42][40]）。

前文第 1 节已介绍点集拓扑。进一步，用多个点和由之框出的凸集即可定义单形（simplex）。单形有真面（proper face）和边缘（boundary）。用众多单形联合而成的集合可定义出复形（complex），过程中要求满足两个条件：单形的真面也在集合中，以及规则相处。

在此基础上即可研究多面体的拓扑：把多面体进行三角剖分（triangulation），形成复形。三角剖分的方式可不限于一种，过程中注意两点：一、需要对单形赋予定向，变成有向单形。作用是保证剖分后相邻两单形的边缘由于定向相反而发生抵消；二、弯曲多面体允许通过连续同胚变换而化为容易剖分的平直多面体来处理。

考虑一个多面体的复形$K$。设$K$中的$p$维单形（由$p+1$个线性独立的点来定义）共有$l$个。则这$l$个$p$-单形的线性组合就成为一条$p$-链，记为$c_p$；最基本的，其组合系数如果取整数，就意味着讨论整数链群。令所有的组合系数各自跑尽整数集$\mathbf{Z}$，就得到所有的$p$-链，集合起来就形成$p$-链群（或称下$p$-链群，$p$-Chain group），记为$C_p$。在$C_p$中我们关心它的两个子群，一个是由所有$p$-闭链形成的闭链群（或称下$p$-闭链群，$p$-Closed chain group），记为$Z_p$；另一个是由所有$p$-边缘链形成的边缘链群（或称下$p$-边缘链群，$p$-Boundary chain group），记为$B_p$。这里所谓的闭链是指该$c_p$是封闭的，自身不再有边缘，即$\partial c_p = 0$，其中$\partial$是边缘算子；边缘链是指该$c_p$自身是某个更高阶$(p+1)$-链的边缘，即$c_p = \partial c'_{p+1}$。容易想到，已经是他人边缘的链是封闭的，自身不可能再有边缘。所以边缘链一定是闭链，对应于事实$\partial^2 = 0$（幂零性，nilpotency）；反之则不然，一条闭链不一定是边缘链，因此$B_p$是$Z_p$的子群。

边缘链总是别人的边缘，因此可缩（contractible），称为是恰当的（exact），数学上认为这种情形平庸；与此形成对比的是那些非边缘的闭链，它们既是封闭的又非他人的边缘，因此不可缩，正是拓扑学要关注的对象。简而言之，拓扑学上有兴趣的关注对象就是"既闭又非恰当"的情形。有鉴于此，特意从$Z_p$中刨除掉$B_p$，就定义出商群

$$H_p = \frac{Z_p}{B_p}, \qquad (40)$$

称为$p$阶同调群（或$p$阶下同调群，Homology group），以此来表征非平庸拓扑。举例：

- □ Möbius 带：$H_0 = H_1 = \mathbf{Z}$（$\mathbf{Z}$是整数集，表示可绕整数圈或覆盖整数次）。一般地，$H_0$含几个$\mathbf{Z}$表示有几个不连通分支；$H_2 = \{0\}$即单点集，表示所有二维覆盖都可缩，平庸。

- 射影平面（Projective plane）$P^2$：$H_0 = \mathbf{Z}$；$H_1 = \mathbf{Z}_2 = \{0,1\}$，二阶循环群；$H_2 = \{0\}$。
- 球$S^2$：$H_0 = \mathbf{Z}$；$H_1 = \{0\}$，即所有一维覆盖都可缩，不存在二维洞，平庸；$H_2 = \mathbf{Z}$，即存在一个三维洞。
- 轮胎面$T^2$：$H_0 = \mathbf{Z}$；$H_1 = \mathbf{Z} \oplus \mathbf{Z}$，表示存在两个二维洞；$H_2 = \mathbf{Z}$，存在一个三维洞。
- Klein 瓶：$H_0 = \mathbf{Z}$；$H_1 = \mathbf{Z} \oplus \mathbf{Z}_2$，那个$\mathbf{Z}_2$标示非定向性；$H_2 = \{0\}$，不存在三维洞。
- 任意二维紧曲面必同胚于如下三种类型之一（紧曲面分类定理）：
  - 球$S^2$。这是定向情况；
  - $n$个$T^2$的连通和（即粘接）。每个$T^2$称为一个环柄（handle）。这是定向情况；
  - $n$个$P^2$的连通和。这是非定向情况。

以上是对底流形拓扑的研究，属于代数拓扑方法；但在应用科学中需要有可操作性，需要有实际可计算的量，最好能够用到微积分。这就需要用上同调的思想对问题进行转化。简言之，上同调就是先把底流形投射到自己的基底上去（基的选择方法有很多，比如上文的$p$-链$c_p$），然后通过寻找这些基的对偶基，把问题转化到对偶空间里去讨论。这里所谓的对偶也有不同的选择，端赖采取什么观点、在何种意义之下的对偶。举简单常见的例子：设流形上存在一个磁矢势$\mathbf{A}$，在外微分语言中是一个 1-形式，$A = A_i dx^i$。现选取两条闭路径$C^I$和$C^{II}$做围道积分，$\frac{1}{2\pi}\oint_{C^I} A^I$和$\frac{1}{2\pi}\oint_{C^{II}} A^{II}$（$A^I$表示是$C^I$上的矢量势，$A^{II}$是$C^{II}$上的势；为简单起见取单位磁通为 1）。那么由 Stokes 定理知道积分的结果取决于围道中是否存在奇点。设$C^I$有奇点，只转一圈，结果就为 1；$C^{II}$没有，结果为零。这样一来，若把积分视为一种特别的内积，就有

$$\langle C^\alpha, A^\beta \rangle = \frac{1}{2\pi}\oint_{C^\alpha} A^\beta = \delta^{\alpha\beta} = \begin{cases} 1, & \alpha = \beta; \\ 0, & \alpha \neq \beta, \end{cases} \quad (41)$$

其中$\alpha, \beta = I, II$。

内积定义出一种对偶关系，对底流形各种链群的讨论就转化为对其上各种外微分形式的讨论。于是就有了上链群$C^p$（"上"表示对偶，Co-chain group）、上闭链群$Z^p$（Co-closed chain group）、上边缘链群$B^p$（Co-boundary chain group）。而对底流形上表征拓扑的各种同调群的研究，就转化为对所谓上同调群$H^p$（Cohomology）的研究，$H^p = \frac{Z^p}{B^p}$。上面的基于外微分形式的上同调，特别被称为 de Rham 上同调，记为$H^p_{dR}$（不需特别说明时也简记为$H^p$）。如前所述，对偶的选择可以有其他方式，那么上同调也可以有其他种类，是基于不同意义而言的工具，适用于不同的问题。

下面的任务就是如何构造拓扑不变的上同调群 — 外微分意义下可称上同调类。构造的思路可采取所谓的陈-Weil 同态（Chern-Weil homomorphism）方法，即先找到满足齐次性、$k$-线性、宗量交换对称性的不变多项式环，然后将当中的宗量代换为流形上的曲率张量，就得到拓扑示性类（topological characteristic class）。重要例子有结构群为$SU(N)$群的复矢量丛上的陈（省身）类，$O(N)$群实矢丛上的 Pontrjagin 类，$SO(N)$群实矢丛上的 Euler 类（复化后即是对应维数上的顶陈类）等等。下面以陈类为例展示一下示性类的结构：

- 总陈类：设纤维丛为$E$，其上的曲率 2-形式为$\Omega$，$\Omega = \frac{1}{2}\Omega^a_{ij} T_a dx^i \wedge dx^j$，其中$T_a$是结构群$G$的生成元，$\Omega_{ij}$在相应的李代数基上展开为矢量。当$G$为阿贝尔的$U(1)$时，$\Omega$可用涡度$\boldsymbol{\omega}$来理解。总陈类定义为

$$C(E) = det\left(I + \frac{i}{2\pi}\Omega\right). \quad (42)$$

行列式展开后是一个无穷求和。由于流形的维数有限，那么由外微分微元的反对易性知高阶的展开项都为零，展开式到某个$m$项就会发生截断（truncation）。于是有

$$C(E) = \sum_{k=0}^{m} C_k(E). \quad (43)$$

- 其中每一项$C_k(E)$都是拓扑不变量，称为第$k$陈类，其积分的结果是整拓扑数，称为第$k$陈数：

$$C_k(E) = \left(\frac{i}{2\pi}\right)^k \sum_{(i)(j)} \delta^{i_1 \cdots i_k, j_1 \cdots j_k} \Omega_{i_1 j_1} \cdots \Omega_{i_k j_k}, \quad (44)$$

其中$\delta^{i_1 \cdots i_k, j_1 \cdots j_k}$是行列式

$$\delta^{i_1 \cdots i_k, j_1 \cdots j_k} = det\begin{pmatrix} \delta^{i_1 j_1} & \cdots & \delta^{i_1 j_m} \\ \vdots & \ddots & \vdots \\ \delta^{i_m j_1} & \cdots & \delta^{i_m j_m} \end{pmatrix}. \quad (45)$$

前几项写出来是：

第一陈类： $C_1(E) = \frac{i}{2\pi} Tr\,\Omega$, (46)

第二陈类： $C_2(E) = \frac{1}{8\pi^2}[Tr\,(\Omega \wedge \Omega) - Tr\,(\Omega) \wedge Tr\,(\Omega)]$, (47)

等等，其中$Tr$是求矩阵迹运算（trace）。从而

$$C(E) = 1 + C_1(E) + C_2(E) + \cdots. \quad (48)$$

示性类的实用性很强，是非常有用的工具；容易看出上面的第一陈类直接给出涡度的积分，是磁通单元的整数倍。该整数就是一个拓扑数，即第一陈数。

接下来的问题是上述的各种示性类都只能取偶维数的微分形式，原因在于其最基本的构造单元曲率$\Omega$是2-形式，那么由曲率组合出来的示性类就只能取2、4、6、…维微分形式。为了解决奇维数流形的拓扑示性问题，陈省身和 Simons 提出了次级示性类（secondary characteristic class）的解决方案，其中最著名的是 Chern-Simons（CS）形式。

把第$k$陈类$C_k(\Omega)$视为包含曲率$\Omega$的$k$次方的多项式$P(\Omega^k)$。$C_k(\Omega)$作为拓扑上有研究价值的示性类，全局上（globally）当然必须是闭而非恰当的2$k$-形式；但是局域地（locally）可以写成一个恰当形式

$$C_k(\Omega) = dQ_{k-1}(a, \Omega), \quad (49)$$

其中$a$是联络 1-形式，$a = a_i^a T_a dx^i$，其协变微商给出曲率$\Omega = da - a \wedge a$。而$Q_{k-1}(a, \Omega)$是

$$Q_{k-1}(a, \Omega) = k\int_0^1 P(a, \Omega_t^{k-1})dt, t \in [0,1]. \quad (50)$$

$P(a, \Omega_t^{k-1})$是说把原来的$P(\Omega^k)$中的一个$\Omega$代换成 1-形式$a$，而其余的$\Omega$代换成新的 2-形式$\Omega_t$，

$$\Omega_t = tda + t^2 a \wedge a, t \in [0,1]. \quad (51)$$

最重要的 CS 形式是 3-形式：

$$CS(a) = a \wedge da + \frac{2}{3}a \wedge a \wedge a = a \wedge \Omega - \frac{1}{3}a \wedge a \wedge a \quad (52)$$

其中第二项（非线性项）的存在是由于结构群$G$的一般情况是非阿贝尔的，故$a$不可对易。但当$G$特别地是可对易的阿贝尔$U(1)$群时， $a \wedge a = 0$，非线性项消失，上式成为

$$CS(a) = a \wedge da. \quad (53)$$

立刻看出，这和前文(26)式对 Helicity 的外微分定义一致。由此可知，螺度实质上具有阿贝尔 Chern-Simons 3-形式的意义。

由非阿贝尔 CS 形式衍生出的 Chern-Simons 场论是最重要的低维拓扑量子场论之一。它与纽结理论、共形场论、顶角算子代数、圈量子引力等重要领域都有关系。特别是，该理论中天然蕴含有各种纽结不变量，在此领域已产生多个数学菲尔兹奖的工作[50]。参图 6-6。

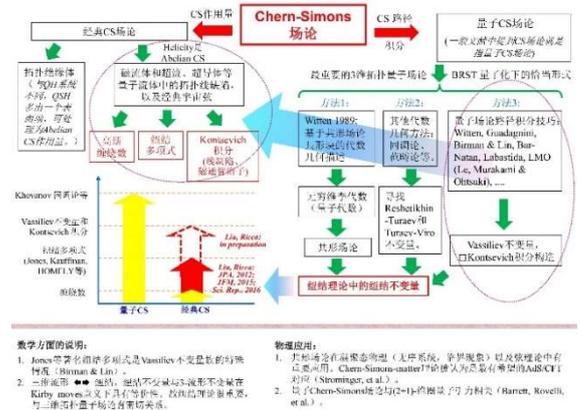

**图 6-6** 陈-Simons 拓扑经典与量子场论与其他领域（特别是纽结理论）之间的关系图。量子场论中的路径积分思想与部分技巧可以应用到拓扑流体力学领域。

**Figure 6-6** The relationship between the Chern-Simons classical and quantum field theories and other research areas, in particular knot theory. The main idea and some techniques of path integrals in quantum field theory can be transplanted to topological fluid mechanics.

经典拓扑流体力学中把理想流动的纽结性质描述清楚本身就是很不平凡的课题，但是力学和物理工作者显然不满足于"说说而已"。如此直观优美图像如何"用"起来而不是束之高阁，是另一个非常有趣和挑战的课题。比如，读者可能就会想到流体纽结能否有可能体现出一些真实纽结的不同捆绑紧固效果，等等。这是可能的：这方面的基础研究已经越来越明确，工程应用研究潜力应该也是巨大的，我们将在最后一节通过指出朱建州及其他合作者在有螺流动方面的相关统计流体力学工作再回来明确这一点。

## 6.3 研究流体纽结拓扑的新工具: 基于 Helicity 构造纽结多项式 (详情咨询: xin.liu@bjut.edu.cn)

量子陈-Simons 场论与各种纽结拓扑不变量的关系给出一个启发：在拓扑流体力学这一经典场论的研究领域中，或许也可以寻找基于 Helicity 的、比（自）缠绕数的拓扑示性能力更强的纽结不变量。以下总结回顾刘鑫与 Ricca 的工作[35][44][36]的主要思路和结果。

事实上，前文所给出的（自）缠绕数在纽结理论中是较弱的纽结拓扑不变量。一个典型的例子是图 6-4 中的三组链环，它们的总高斯互缠绕数都为零，所以仅凭互缠绕无法区分它们。另一个著名的例子是图 6-7 中的平庸环和 8-字结，其自缠绕数都等于零，故无法仅凭自缠绕数区分它们。

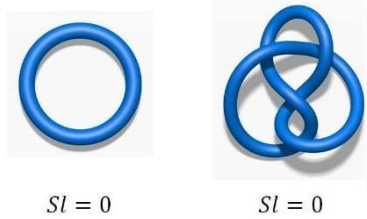

**图 6-7** 平庸环（左）和 8-字结（右）。平庸环显然自缠绕数为零；8-字结的自缠绕数可计算也等于零。
**Figure 6-7** A trivial circle (*left*) and a figure-8 knot (*right*). The self-linking number of the trivial circle is obviously zero; that of the figure-8 knot vanishes as well, as can be easily verified.

因此就需要在流体纽结拓扑的研究中引进示性能力更强的纽结不变量。

另一方面，拓扑量子场论中的 CS 路径积分定义为 Wilson loop 的真空期望值（对应于数学中的和乐群）：

$$\langle \prod_i e^{i\oint_{\gamma_i} a} \rangle = \frac{1}{Z} \int [Da] \left( e^{i\Sigma_i \oint_{\gamma_i} a} \right) e^{iCS(a)}, \quad (54)$$

其中[Da]是积分测度，表示积分在联络$a$（规范场论的术语也叫规范势）的空间进行；$CS(a)$是前文(52)式给出的 CS 作用量；$\gamma_i$表示第$i$个纽结；$e^{i\Sigma_i \oint_{\gamma_i} a}$称为 Wilson loop。为简单起见，上式中未进行 BRST 量子化手续，也即未加入规范固定项（gauge fixing）和鬼场项（ghosts）。拓扑场论中的著名工作[50]指出这个积分可以给出纽结多项式拓扑不变量。

这为分析 Helicity 的拓扑内涵提供了线索。前(53)式指出它是一个阿贝尔的 CS 作用量，也即此时(54)式右边的被积函数变成了

$$e^{i\Sigma_i \oint_{\gamma_i} u} e^{i \int u \wedge du}, \quad (55)$$

其中联络$a$已写成速度场$u$（外微分中的 1-形式）。作用量即$H = CS(u) = \int_V u \wedge du$。进一步注意到，当考虑极细涡丝（thin vortex filament）— 磁流体力学中考虑极细磁通管 — 的时候，涡度场可认为沿着涡丝的方向，$\omega = \omega_0 \hat{t}$，其中$\omega_0$为常数，$\hat{t}$是涡丝中轴线单位切矢。在此条件下，$H$就可写到涡丝线积分上去[23]

$$H = \sum_i \Phi_i \oint_{\gamma_i} u = \sum_i \oint_{\gamma_i} u \text{ 从而} e^{iH} = e^{i\Sigma_i \oint_{\gamma_i} u}, \quad (56)$$

其中每根涡丝的$\Phi_i$（circulation）为简单起见已取为 1。这样一来，发现(55)式中的两个因子变成相同。这就表明，是这个指数形式$e^{iH} = e^{i\Sigma_i \oint_{\gamma_i} u}$具备提供纽结多项式不变量的能力。

基于这一认识，就可以抛却泛函路径积分体系，一切从研究这个指数形式出发。下面将证明，该形式可以给出超出（自）缠绕数的更高阶的拓扑信息，如纽结多项式等。在给出证明之前，先就可能产生的疑问提供几点说明：

☐ 目前国际上对螺度的研究，已有多个理论和实验组转而开始关注线积分形式的 Helicity（称为 Field line helicity），如邓迪大学太阳等离子体物理团队[24][27]。

☐ 指数形式可以退回到非指数的形式。粗略的理解是，如果对该形式进行指数展开，那么一阶项将会回到通常的情形，描述（自）缠绕数这样的低阶拓扑不变量—实质上，这意味着只考虑每根纽结都拓扑守恒的情形，不考虑它们之间的相互作用；而二阶以上的项，则会包含多个线积分，那就必须考虑纽结片段相碰时发生破坏拓扑守恒的各种相互作用的情况，如重联（reconnection）。考察指数形式，实际上意味着寻回之前扔掉的高阶项。

- 下文中可以看到，指数形式与线积分搭配使用有以下优点：
  - 由于积分路径是曲线，所以允许对路径进行断开、重联，以及添加和删减虚拟路径等操作。所得的结果在指数上具有可加性，写在指数下则是相乘因子。即 $e^{\int_L u} = e^{\int_{L_1+L_2} u} = e^{\int_{L_1} u} e^{\int_{L_2} u}$，其中路径具有可加性

$$L = L_1 + L_2。 \tag{57}$$

  - 这就允许把上面路径操作之后所得的某些新因子进行具体计算，得到常数。这些常数即可定义为纽结多项式的定义式（即拆接关系，skein relation）所需要的那些常数。

由此可以证明，上述指数形式可以给出 Jones、HOMFLYPT 等纽结多项式拓扑不变量。下面对构造/证明思路给出简短说明。

一个纽结（不计粗细的）是指三维当中的一条封闭曲线，数学上表达为一个映射 $\gamma: S^1 \to \mathbf{R}^3$。映射的方式不同意味着打结的方式不同，有平庸的无缠绕环，也有非平庸的结。把纽结画成平面图，纽结和纽结之间的形态不同就体现在各个交点处的交叉方式不同上面，如图 6-8。

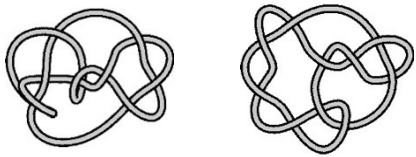

图 6-8 两个 10 交叉点纽结，10-22（左）和 10-35（右）。二者有相同的 Jones 多项式，但 HOMFLYPT 多项式不同。

**Figure 6-8** Two knots with 10 crossing sites: 10-22 (*left*) and 10-35 (*right*). They share the same Jones polynomial, but have different HOMFLYPT polynomials.

进一步地，多个纽结同时出现，相互之间可以有嵌套或无嵌套。这样形成的组合称为链环（link），当中的每个纽结称为链环的一个分量（component）。显然，单个纽结是链环的特殊情况。一般说的纽结拓扑不变量（多项式等）更准确地说应叫做链环不变量，因为它们处理的是更一般的链环的拓扑示性问题。

两个链环画到平面上以后形态不同，并不意味着它们相互不等价——它们之间可能通过连续拓扑变换从一个变到另一个。如果这点能做到，就称它们拓扑等价或同痕（isotopic），否则为不等价。纽结理论的核心任务之一即是分辨两个纽结是否等价；若不等价，相差多少。为完成这一任务数学上发展出了纽结拓扑不变量方法，即寻找一些在拓扑变换下不变的量或不变的性质，以此判断两个纽结等价与否。注意，拓扑不变量是必要而非充分条件，也即：若某拓扑不变量发生了变化，立可判断二者不等价；若未发生变化，只能说在该不变量之下等价。若有更高阶的不变量发生了变化，仍判断此二者为不等价。

那么，找寻拓扑不变量需要先分析连续拓扑变化；Reidemeister 指出，所有的变化归根结蒂都由以下三种最基本的变化组合而成，称为 Reidemeister moves。如图 6-9。

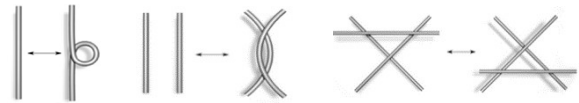

图 6-9 三种 Reidemeister moves：（左）Type-I；（中）Type-II；（右）Type-III。 纽结拓扑不变量，三种变换下都不变的，称为 ambient isotopic；只在后两种之下不变的，称为 regular isotopic。

**Figure 6-9** Three types of Reidemeister moves: Type-I (*left*), Type-II (*middle*) and Type-III (*right*). A knot topological quantity which is invariant under all these three types of moves is called an ambient isotopic invariant; a quantity which is invariant under the latter two types of moves is called a regular isotopic invariant.

接下来，按数学上的思维方式，考察纽结时专注于每个交叉点。对一个具体的交叉点来说，无非是三种交叉方式，如图 6-10。

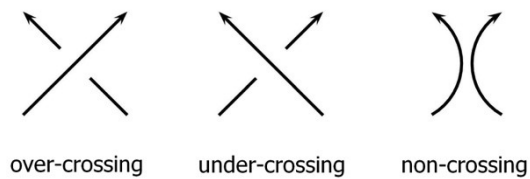

图 6-10 考察三个链环，它们几乎处处相同，只在一个交

叉点处不同：（左）上交叉，over-crossing；（中）下交叉，under-crossing；（右）不交叉，non-crossing。

**Figure 6-10** Three links, which are almost the same everywhere, except one particular crossing site: over-crossing (*left*), under-crossing (*middle*) and non-crossing (*right*).

于是，所谓的纽结多项式的定义就是针对这三种交叉方式之间相互关系的公式（称为拆接关系，skein relations）；某种意义上说，可认为是一种递推关系[31]。如：

☐ Jones 多项式 skein relations，关于单参数$\tau$[29]：

$$V(\text{circle}) = 1, \quad (58)$$

$$\tau^{-1}V(\text{over}) - \tau V(\text{under}) = \left(\tau^{\frac{1}{2}} - \tau^{-\frac{1}{2}}\right)V(\text{non}). \quad (59)$$

☐ HOMFLYPT 多项式 skein relations，关于双参数$a$和$z$[27][43]：

$$P(\text{circle}) = 1, \quad (60)$$

$$aP(\text{over}) - a^{-1}V(\text{under}) = zV(\text{non}). \quad (61)$$

公式中的"circle"指平庸圆，"over"指上图中的 over-crossing 态，"under"指 under-crossing，"non" 指 non-crossing 态。

所谓的构造纽结多项式，就是构造上面各种拆接关系中的参数$\tau$、$a$、$z$等，并证明它们满足那些拆接关系。

构造 Jones 多项式的拆接关系(58)和(59)需要构造参数$\tau$。文献[35][44]发展出了一个所谓的"添加/删除虚拟路径"技巧，如图 6-11。

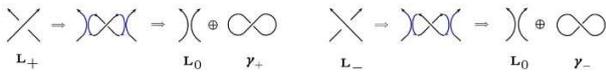

**图 6-11** 添加/删除虚拟路径：(a)对 over-crossing（记作$L_+$）的处理。通过添加虚拟路径（蓝色；由于定向不同，相互可以抵消），$L_+$被化为一个 non-crossing（记作$L_0$）加上一个小的 unknot（记作$\gamma_+$）；(b) 对 under-crossing（记作$L_-$）的处理。$L_-$被化为一个$L_0$加上一个小的 unknot（记作$\gamma_-$）。$\gamma_+$带一个正向的$\frac{1}{2}$拧转，$\gamma_-$带一个负向的$\frac{1}{2}$拧转。

**Figure 6-11** Technique of adding/subtracting imaginary paths: (a) The treatment for an over-crossing site, denoted as $L_+$: by adding imaginary paths denoted by the blue color (which are able to cancel each other due to opposite orientations), $L_+$ is turned into a non-crossing, denoted as $L_0$, plus a small unknot, denoted as $\gamma_+$; (b) That for an under-crossing site, denoted as $L_-$: $L_-$ is turned into an $L_0$ plus a small unknot denoted as $\gamma_-$. The $\gamma_+$ carries a positive twist, $\frac{1}{2}$; the $\gamma_-$ carries a negative twist, $\frac{1}{2}$.

于是，如果以图中的$L_+$、$L_-$、$L_0$、$\gamma_+$和$\gamma_-$为线积分的积分路径，就有

$$e^{\oint_{L_+}} = e^{\oint_{L_0 \oplus \gamma_+}} = e^{\oint_{\gamma_+}} e^{\oint_{L_0}} = ke^{\oint_{L_0}}, \quad (62)$$

$$e^{\oint_{L_-}} = e^{\oint_{L_0 \oplus \gamma_-}} = e^{\oint_{\gamma_-}} e^{\oint_{L_0}} = k^{-1}e^{\oint_{L_0}}, \quad (63)$$

其中定义$k = e^{\oint_{\gamma_+}}$；可计算恰好有$k^{-1} = e^{\oint_{\gamma_-}}$。那么，如果定义$\langle L_+ \rangle = e^{\oint_{L_+}}$、$\langle L_- \rangle = e^{\oint_{L_-}}$和$\langle L_0 \rangle = e^{\oint_{L_0}}$，就得到一组原始的关系：

$$\langle L_+ \rangle = k\langle L_0 \rangle, \quad \langle L_- \rangle = k^{-1}\langle L_0 \rangle. \quad (64)$$

这已初具模样，接近最终所需拆接关系。接下来，经过一系列步骤（先去掉纽结的定向，用各态遍历假定得到 Kauffman 尖括号多项式的拆接关系[30]，然后再重新赋予纽结定向），即可得到 Jones 多项式的拆接关系(58)和(59)式。具体参文献[35]。

上面的 Jones 多项式是单参数多项式。示性能力虽然比（自）缠绕数强许多，但仍然有局限性，例如它无法分辨前文图 6-8 中的 10-22 和 10-35 那两个纽结。为此文献[27][43]构造了一种推广的（generalized）Jones 多项式，即 HOMFLYPT 多项式（"HOMFLYPT" 是两篇论文共八位作者姓氏的首字母）。后者是双参数多项式，如(60)和(61)式所示，因此拓扑示性能力更强，可以分辨图 6-8 中的 10-22 和 10-35 纽结。

在 HOMFLYPT 多项式拆接关系 60)和(61)式的构造中，除了要用到上面"添加/删除虚拟路径"的技巧来构造参数$z$以外，还要用到 Dehn surgery 技术来构造参数$a$，如图 6-12。$z$的构造过程主要体现了 writhing 的贡献，而$a$则体现 twisting 的贡献，参前文(37)式。HOMFLYPT 多项式的具体构造细节参文献[36]。

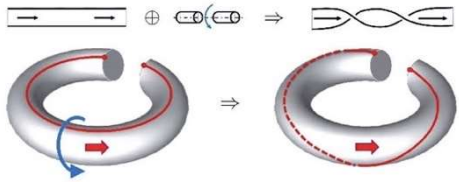

**图 6-12** Dehn surgery 技术。用于构造 HOMFLYPT 多项式拆接关系中的参数$a$。

**Figure 6-12** The technique of Dehn surgery. It is employed to construct the parameter $a$ in the skein relations of the HOMFLYPT polynomial.

流体纽结的多项式拓扑不变量方法，不仅可用于拓扑示性，还可以有一个重要的用途即对流体纽结的拓扑复杂度进行分层管理。自然界中有许多的物理纽结链环，其拓扑复杂度级联递降过程往往伴随着能量、熵等动力学性质的改变。比如：

- 太阳等离子体[47]：适用磁流体力学。太阳表面的日冕物质抛射（coronal mass ejections）会伴随大量的放能现象，很重要的一个原因是大量磁通管所形成的链环发生磁重联。这是拓扑非守恒变化，能量降低的同时伴随拓扑系综的复杂度降低。

- 水中实现的纽结状涡旋管[33][34]：其结构不能长期保持，会发生拓扑非守恒的退化。其过程是经历一系列的逐级退化 ——不是一步到位，从复杂的纠缠状态一次性崩裂成平庸碎片— 每一步只经历一个重联事件，每一步是非平衡过程中的一个亚稳态。最终产物是众多平庸的微小涡圈。如图 6-13。

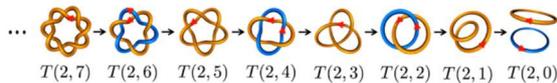

**图 6-13** 环面纽结/链环$T(2,n)$的级联递降退化过程。
**Figure 6-13** Cascade degeneration procedure of a torus knot/link $T(2,n)$.

- 生物中 DNA 质体的重组[48][49]： 退化的过程与上图水中的纽结涡旋管退化非常相似。单数值试验表明存在多种途径，各自发生的概率并不均衡，有大有小。如图 6-14。

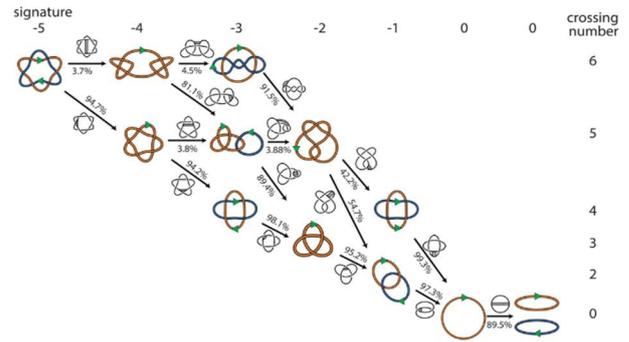

**图 6-14** DNA 环面链环$T(2,6)$的级联退化过程。存在多条退化途径，数值试验给出了各路径发生的概率。
**Figure 6-14** Cascade degeneration procedure of a DNA torus link T(2,6). There exist various routes of degeneration, which have different probabilities of occurrence according to numerical simulations.

为了与物理纽结的实验观察相比较，刘鑫与 Ricca 在工作[37][45]中对 HOMFLYPT 多项式中的参数$a$和$z$进行了赋值，所采用的数值来源于构造$a$、$z$的时候物理过程的具体含义。然后计算环面纽结/链环$T(2,n)$的 HOMFLYPT 多项式的具体数值，发现随着$n$的降低该多项式数值也单调降低，符合上述级联退化的特征。

**图 6-15** 环面纽结链环的 HOMFLYPT 多项式的数值，形成单调递降级联序列。

**Figure 6-15** Numerical values of the HOMFLYPT polynomials of torus knot/links, which form a monotonically degenerating cascade.

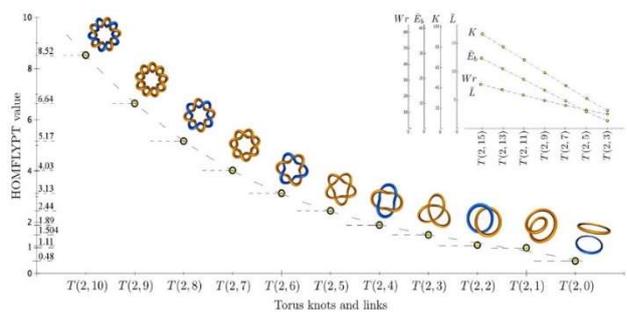

## 7 引入内结构和滑移粘性的流动规范场唯象理论(详情咨询:zouwn@ncu.edu.cn)

### 7.1 运动流体作为时空流形的建模

首先从一般连续介质理论的角度展开讨论，以便

对比分析。介质（比如常见的水、空气、金属材料和混凝土等）的连续性假设是建立固体变形和流体流动的宏观唯象理论的基础。在连续性假设下，经典连续介质模型[52]认为作为连续介质的物体由无数的物质点（作为用于统计该点的物理量的所有微粒的代表）连续聚集而成、不同时刻的物体存在（质点间的）一一对应关系。在经典模型中，物体完全用它占据的坐标域表征，这又称为物体的构型。构型是质点的连续统，物体的内部变化则是同胚演化，即如果初始静止状态下的物体同胚于欧氏空间，则后续任何时刻的物体也都同胚于欧氏空间。经典连续介质模型中的质点没有尺度、只有位置的变化，但它与牛顿力学中的质点仍有些不同（参见[53]，为示区别，下面暂称前者为物质点），比如：(1)物质点有密度而无质量，(2)即便忽略微粒振动和扩散因素，物质点所包含的分子（原子）一般不是确定的，固体弹性变形的物质点的内含具有不变性，但流体流动的物质点的内含必然是变化的，(3)必须考虑分子（原子）甚至晶粒尺度的微结构特性（比如液晶、固体晶体等）时，还不得不有所突破经典模型，采用广义连续介质模型赋予物质点以旋转、变形等几何特征[54]。由此，在经典连续介质模型中，固体的可恢复的弹性变形和不可恢复的塑性变形、流体的流动等在几何上是不加区分的，它们只在本构关系中才能加以分别。

从历史发展来看，经典连续介质模型源自弹性理论，比如 Maxwell 曾称流体粘性为瞬时弹性（fugitive elasticity），并且物质点常常借用质点的封闭孤立的特征，在几何分析中被标示和跟踪，更有等价的所谓拉格朗日（物质）描述和欧拉（空间）描述。由于物质点没有尺度特征，可以对介质作无限分割，但又不能分割到分子（原子）尺度，因此常有"物质点是宏观无穷小、微观无穷大"的说法。对某些材料，广义连续介质模型具有物理上的必要性，但赋予物质点旋转

或一般变形特征后要考虑它的转动惯量必然又会导致模型的不自洽。如此看来，经典连续介质模型虽然在数学上具有自恰性，但物理上并不真实[55] [56]。

广义连续介质模型的研究中很少去思考介质变化过程中自发形成的内结构[57]。我们曾对于固体提出一种唯像模型[58]：对于一般弹性变形，**固体的流形描述是用变形度量定义的黎曼流形**，而理想弹性的变形协调条件意味着这一黎曼流形具有处处为零的曲率。对于流体流动提出的模型则强调伴随着分子尺度的层状序化的内部滑移，由此层状序化方向和滑移方向赋予了流体元一种（内部）取向结构，从而可以通过沿不同方向的旋转来定义流体元的同构关系的；更进一步，这一旋转的不可积性给出了流体整体性的拓扑结构。由此**流体的流形描述是用取向联络定义的微分流形**。

通过考虑取向联络通过规范速度场的微分进入粘性力的本构关系，以及小尺度涡旋对取向的影响，流形描述下流体粘性的各种力矩作用也就由此粉墨登场。

## 7.2 涡旋场作为规范场的滑移粘性流动理论

以下完全回到粘性流体的流动问题。根据前述模型，考虑到流体不能抗拒剪切，加上不可压流动中流体的物质疏密度在宏观意义上的不变，流体的序化就只能是滑移序化，从而表达局部同构的一般线性变换群只能是旋转群。由此，流动分析的最小单元不再假设为孤立的流体质点，而是具有微尺度的、可以互相交叠的流体元（图 7-1），一个流体元就是局域时空中的一团有取向特征的流体分子。流体元的取向没有绝对的意义，但相邻流体元的取向差定义了流体元的同构平移，即比较不同流体元速度的平行移动要伴随一定的旋转。

描述流体元状态和相互关联的模型表述如下[9]：

---

[9] 需要指出的是，此类模型是流体力学研究史上从未有过的[59]。流动理论是最早的场理论，流动涡旋图像曾是 Maxwell 研究电磁理论模型的模板，曾几何时，如上一节所示，磁链环的理论研究成果却成了流动拓扑学研究的样版。现已清楚，电磁场矢量应该称作电磁场强，相应的电磁势是带电粒子波函数场的联络（或称规范场，见脚注 12）。另一方面，把粘性力看作流体内部的摩擦力是很早就固化的一个直觉，牛顿提出了滑移粘性的模型，圣维南曾试图给出这个模型的一般数学表达而未成功，本文模型可以看作是这种努力的一个延续。显而易见的是，这需要微分流形这一新的数学工具，当了解到电磁场理论以及各种现代场理论都得益于这一工具而取得进展和成功，我们有信心顺此发展流动的新型场理论。

在流体内一点的流体元对它包含的流体微粒的运动和序化状态进行统计平均，不仅有速度 **V**、还有反映流体元时空取向变化的涡旋场 **W**，后者包括自旋场 **Φ** 和弯扭场 **A**，它们可以组合表示成轴矢量值的微分 1-形式，即[10]

$$W^i = A^i_k dx_k + \Phi^i dt。 \quad (65)$$

这样引进的涡旋场的核心意义有两点：1）粘性作用首先取决于流体运动的不均匀性，而速度场不均匀性的计算需要考虑运动流体的滑移结构（它反映了流体的拓扑形态，数学上对应于微分流形上的联络结构），即速度场作为流形上的矢量丛，其不均匀性引起的粘性作用时必须考虑联络结构的协同效应，从而在粘性作用表达中须采用如下的协变导数[11]

$$D_k V_i = \partial_k V_i + e_{ilm} A^l_k V_m； \quad (66)$$

2）建立了小尺度涡旋的序参量 **Φ** 描述及其与流体元取向结构的关联。假设流体元中的每个流体分子除了位置外都还有一个取向特征可以标示并且可以短时跟踪的话，当前时刻流体元的所有流体分子的取向最终有一个平均取向，单位时间内所有流体分子的位置改变、取向改变的平均就定义了速度和自旋（小尺度涡旋的序化参量）。这里原则上流体元也可以是整体旋转，但在湍流中未必真有这种整体取向改变的发生。

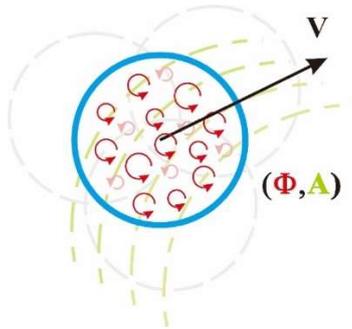

图 7-1 流体元模型
**Figure 7-1** Model of fluid element.

连续介质理论中作为标量的能量（能量变化率）是赋予体元（或质量元，即密度乘体元）的，所以广义力和广义流的微分形式表达互相共轭，它们的外积得到体形式。举例来说，我们常说应力是二阶张量，它的两个方向指标一个表示力、一个表示所作用的微面，当观察者旋转时两个方向一起随之协变。但是力的方向和面的方向是完全不同的特征，应力的客观性应该是指作用在同一个微面上的应力会随之观察者协变，而作用面对应的指标是固定对应于某个微面上的，而"微面上的集度"这种密度概念用微分形式表达是更自然的。这样，如同文献中已经有学者指出的 [61]，应该把应力写成矢量值的 2-形式 $\boldsymbol{\sigma} = \sigma_{ij} \mathbf{e}_i da_j$（或时空 3-形式 $\boldsymbol{\sigma} = \sigma_{ij} \mathbf{e}_i da_j \wedge dt$，加上 $dt$ 将赋予它力冲量的含义，并便于平衡分析中的运算）[12]。粘性力的起源是流动速度的不均匀，在有滑移结构的背景下这就是速度的协变微分，如（66）式所示。于是，各向同性线性粘性本构关系就可以写成如下紧凑的形式：

$$\boldsymbol{\sigma} = \mu \star DV。 \quad (67)$$

在平直流动（**W** 处处等于零）中，它与教科书中常用的牛顿摩擦力公式也是一致的。

模型中用轴矢量值时空微分 1-形式表示的涡旋场是运动流体内部序化结构的表征，远比速度标量值空间 1-形式来的复杂且重要，后者只是揭示流体宏观速度场结构的一个出发点，前者则是描述滑移粘性作用和湍流多尺度演化的基石。它们（涡旋场-速度场）在数学物理上的差别：时空-空间，联络结构-状态量，非张量特性-矢量特性，旋转-平动，流体元内部序化

---

[10] 以下公式均在笛卡尔坐标系中展开，不再区分协逆变指标，重复指标即表示求和；而用上指标表示轴矢量方向，下指标表示一般矢量方向。
[11] 作者[60]曾从客观性要求的角度讨论在寻找粘性力的运动学起源中采用协变导数的必要性。粘性力是客观存在于流体中的相互作用，在各处不同旋转的观察者看确定的截面上的粘性力应该具有协变性，速度亦应如是。但粘性力源于速度的不均匀性（即速度梯度），而常规的速度梯度是不满足协变性的，由此可以引出弯扭场作为规范场的存在。进一步，流体的动力学方程也应该具有协变性（或形式不变性），弯

扭场也就继续会在粘性力的合成中起作用，此不详述。
[12] 应力作为二阶张量的数学表达是建立现代弹性理论（经典连续介质理论的最佳典范）的里程碑，它刮起的旋风把同样是描述介质内部相互作用的粘性力给带进去了，甚至长期以来让人们对弹性变形和粘性流动在微观机理上的显著不同变得无感。采用矢量值的面积形式表达粘性力会提醒我们提出这样的图像：粘性流体中存在那些特别的（滑移）面，它们定义的（取向）内结构能对流动过程中的滑移摩擦加以规范。

-流体元平动，等等。速度场只有一个指向加一个速率，涡旋场首先是分成时空两部分：时间部分自旋场有一个转轴加一个角速度，空间部分弯扭场则有三个正交指向、每个指向上各有一个转轴加一个转角、三个转轴也互相正交（图 7-2）。

在新模型中，所谓层流是指流体元的取向具有全局可积性的定常流动，由此涡旋场可以用速度场来表示，同时层流中不会有小尺度涡旋运动。一般情况下，涡旋场是独立于速度场的，涡旋场强

$$\mathbf{F}^i = D\mathbf{W}^i = d\mathbf{W}^i + \varepsilon_{ijk}\mathbf{W}^j \wedge \mathbf{W}^k = B_k^i da_k + H_k^i dx_k \wedge dt \quad (68)$$

包括弯扭场强 $B_k^i = \varepsilon_{klm}\left(\partial_l A_m^i + \frac{1}{2}\varepsilon_{ipq}A_l^p A_m^q\right)$ 和自旋场强 $H_k^i = \partial_k \Phi^i - \partial_t A_k^i + \varepsilon_{ipq}A_k^p \Phi^q$ 两部分，反映了涡旋场的拓扑特征，数学上又称为曲率张量。从上述定义，可以导出结构方程（数学上又称为 Bianchi 恒等式）

$$D\mathbf{F}^i = d\mathbf{F}^i + \varepsilon_{ijk}\mathbf{W}^j \wedge \mathbf{F}^k = \mathbf{0} \quad (69)$$

涡旋场有 12 个分量，涡旋场强有 18 个分量，但是 Bianchi 恒等式包含 12 个关系，也就是说，涡旋场强只有 6 个独立场量。实际上，由包含三个独立参量的任意旋转张量给出的可积涡旋场不会产生涡旋场强，这样涡旋场还有 9 个独立的分量，如要与涡旋场强的独立分量对等，仍需要三个自然约束（详见后面的协变守恒关系）。

利用作用量变分原理，从拉氏量密度

$$L_{eu}[V_i] = V_i(\rho a_i - f_i + \partial_i p) \quad (70)$$

可以导出理想不可压流动的欧拉方程

$$M_i = \frac{\partial L_{eu}}{\partial V_i} = \rho a_i - f_i + \partial_i p = 0 \quad (71)$$

其中加速度 $a_i$、外部体积力 $f_i$ 和压力 $p$ 都是在变分处理中保持不变的量。真实流动必须考虑粘性，在 N-S 方程理论中就是增加了粘性项

$$L_{vis}\left[\partial_k V^i\right] = \frac{1}{2}\mu\left(\partial_k V^i\right)\left(\partial_k V^i\right)$$

在新模型中，由于涡旋场的引入[13]，根据规范场论[1][14]的最小置换与最小耦合原理[9][15]，我们构造了如下粘性作用项

$$L_{vis}\left[V_i, W_\mu^j\right] = L_Y[Y_{ki}] + L_\chi[\chi_{ki}] - L_\Phi[\Phi^i] - L_H\left[H_k^i\right], \quad (72)$$

其中 $L_Y, L_\chi, L_\Phi, L_H$ 是广义流 $Y_{ki} = \partial_k V_i + \varepsilon_{ilm}A_k^l V_m$，$\chi_{ki} = \varepsilon_{ilm}B_k^l V_m$，$\Phi^i$，$H_k^i$ 的非负能量泛函。在上述构造中，考虑到小尺度涡旋运动是独立于主流且本性耗散的，能量项 $L_\Phi$ 和 $L_H$ 之前的负号表示粘性作用量在小尺度涡旋和主流之间的互相传递性，而弯扭结构场就起到桥梁作用。这样，基于拉氏量密度

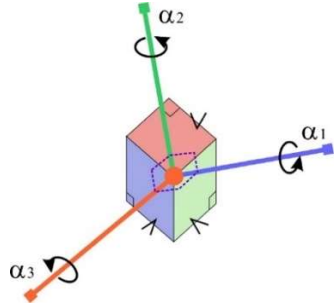

图 7-2 弯扭场的图像

---

[13] 涡旋场之于速度场，犹如电磁理论中的电磁势之于带电粒子波函数，是一种规范场。不同的是，流体中没有常在的带电粒子这样的东西。

[14] 在物理学的研究中基本物理规律所包含的对称性起着非常重要的作用。对称性分为两大类：一类是时空对称性，它们是与描述物理事件的时空坐标的变换（例如时空坐标的平移和Lorentz变换）相联系的；另一类对称性是内部对称性。在场论中，它们是与不改变时空坐标的场的变换相联系的。这种变换称为内部空间的变换。物理学中的变换构成变换群。物理规律的对称性归结为基本方程在这些变换群下的不变性。按照Noether定理，相应于对称群的每个生成元有物理系统的一个守恒律。在场论中可以对不同时空点的场做独立的变换，相应的群元素是时空坐标的函数，这种变换称为定域规范变换，常简称为规范变换[62]。

由于流体中的障碍物和涡旋的演化，流体的底流形通常不能同胚于一个欧氏空间，而需要用一系列同胚欧氏空间的子域来覆盖，子域之间的转换函数构成了旋转群；另一方面，将流体单元的速度看作团聚于此的流体分子的一个统计状态，即即定义了流体底流形上一点的内部空间，速度场就是流形上切丛的一个截面。内部空间的规范变换也构成旋转群，不同点上的内部空间可以通过旋转实现同构，涡旋场就是同构联络。与切丛相配的标架丛是主丛，涡旋场作为主丛上的联络，由于底流形的结构群、标架丛的纤维和对称群都是旋转群，涡旋场就是规范场。

[15] 从 $L_{vis}[\partial_k V_i]$ 到 $L_Y[Y_{ki}]$ 称为最小置换，再到 $L_{vis}[V_i, W_\mu^j]$ 称为最小耦合。最小置换就是用规范协变导数置换没有考虑规范场的一般空间导数；最小耦合就是在拉氏量构造中对独立的规范不变项（过程）只考虑其自耦合，而不考虑交叉耦合。规范场理论容许拉氏量的构造。拉氏量在规范变换下保持不变，由此必然引入传递相互作用的规范场，最小置换与最小耦合原理是构造规范不变的拉氏量的基本准则。

**Figure 7-2** Sketch of swirl field.

$$L\left[V_i, W_\mu^j\right] = L_{eu}\left[V_i\right] + L_{vis}\left[V_i, W_\mu^j\right], \quad (73)$$

可以列出流动过程中与粘性作用相关的各种广义流和广义力如表 7-1，其中有些反映的是本构关系有些则是诱导的耦合作用，比如 $\Sigma_i = \frac{\partial L_\chi}{\partial V_i} = \varepsilon_{ilm} B_k^m \frac{\partial L_\chi}{\partial \chi_{kl}}$。

表 7-1  粘性作用中互为功共轭的广义流和广义力

| 广义流 | 广义力 | 广义力的函数性质 |
| --- | --- | --- |
| $V_i$ | $\Sigma_i = \frac{\partial L_\chi}{\partial V_i} = \varepsilon_{ilm} \Pi_{kl} B_k^m$ | 耦合，矢量值时空体积 4-形式 |
| $\Phi^i$ | $m^i = \frac{\partial L_\Phi}{\partial \Phi^i}$ | 本构，轴矢量值体积 3-形式 |
| $A_k^i$ | $J_k^i = -\frac{\partial L_Y}{\partial A_k^i} = -\varepsilon_{ilm} V_l \sigma_{km}$ | 耦合，轴矢量值时空面积 3-形式 |
| $Y_{ki} = \partial_k V_i + \varepsilon_{ilm} A_k^l V_m$ | $\sigma_{ki} = \frac{\partial L_Y}{\partial Y_{ki}}$ | 本构，矢量值时空面积 3-形式 |
| $\chi_{ki} = \varepsilon_{ilm} B_k^l V_m$ | $\Pi_{ki} = \frac{\partial L_\chi}{\partial \chi_{ki}}$ | 本构，矢量值时空 2-形式 |
| $H_k^i = \partial_k \Phi^i - \partial_t A_k^i + \varepsilon_{ipq} A_k^p \Phi^q$ | $E_k^i = \frac{\partial L_H}{\partial H_k^i}$ | 本构，轴矢量值面积 2-形式 |
| $B_k^i = \varepsilon_{klm}\left(\partial_l A_m^i + \frac{1}{2}\varepsilon_{ipq} A_l^p A_m^q\right)$ | $G_k^i = \frac{\partial L_\chi}{\partial B_k^i} = \varepsilon_{ilm} V_l \Pi_{km}$ | 耦合，轴矢量值时空 2-形式 |

设 $\mathcal{D}$ 是流体占据的时空域，考虑流场的独立性及其在边界上的变分为零，流体系统的最小作用量原理

$$\delta \int_\mathcal{D} L[V_i; A_k^i, \Phi^i] dv \wedge dt = \int_\mathcal{D} \delta L[V_i; A_k^i, \Phi^i] dv \wedge dt = 0 \quad (74)$$

导出广义力的动力学方程如下：

$$\mathbf{M}_i + \mathbf{\Sigma}_i = d\boldsymbol{\sigma}_i + \varepsilon_{ilm} \mathbf{A}^l \wedge \boldsymbol{\sigma}_m = D\boldsymbol{\sigma}_i, \quad (75)$$

$$\mathbf{J}^i = d\mathbf{Q}^i + \varepsilon_{ipq} \mathbf{W}^p \wedge \mathbf{Q}^q = D\mathbf{Q}^i, \quad (76)$$

其中

$$\begin{aligned}\mathbf{J}^i &= \frac{\partial L_\Phi}{\partial \Phi^i} dv - \frac{\partial L_Y}{\partial A_k^i} da_k \wedge dt = \mathbf{m}^i + \varepsilon_{ilm} \boldsymbol{\sigma}_l \wedge \mathbf{u}_m, \\ \mathbf{Q}^i &= \frac{\partial L_H}{\partial H_k^i} da_k - \frac{\partial L_\chi}{\partial B_k^i} dx_k \wedge dt = \mathbf{E}^i - \varepsilon_{ilm} V_l \mathbf{\Pi}_m.\end{aligned} \quad (77)$$

方程(76)包含两个部分：一个是三维空间上的微力矩平衡，一个是通过面积域的微力矩冲量的平衡。这两部分平衡不是完全独立的，存在四维时空域上的高阶平衡

$$\begin{aligned}D\mathbf{J}^i &= d\mathbf{J}^i + \varepsilon_{ipq} \mathbf{W}^p \wedge \mathbf{J}^q = \varepsilon_{ipq} \mathbf{F}^p \wedge \mathbf{Q}^q \\ &= \varepsilon_{ipq}\left(H_k^p E_k^q + B_k^p G_k^q\right) dv \wedge dt.\end{aligned} \quad (78)$$

考虑到广义力和广义流（$\mathbf{H}$ 和 $\mathbf{E}$, $\mathbf{B}$ 和 $\mathbf{G}$）的同轴性，它可以简单是微力矩时空流的协变守恒关系

$$D\mathbf{J}^i = 0。 \quad (79)$$

进一步利用 $\mathbf{m}$ 和 $\boldsymbol{\Phi}$、$\boldsymbol{\sigma}$ 和 $\mathbf{Y}$ 的同轴性，守恒关系(79)可以写成

$$\partial_t m^i = \varepsilon_{ipq} V_p D_k \sigma_{kq}, \quad (80)$$

意味着从粘性（面）力集成的体力与速度的不平行性搓出了对应于小尺度涡旋运动的微力矩。

不同的流体具有不同的本构关系。对于大部分小分子流体，内在的各向同性将保证本构关系中广义力和广义流的同轴性。进一步对于牛顿流体，比如水和空气，更有对应于各向同性线性本构的二次型拉氏量密度

$$\begin{aligned}L\left[V_i, W_\mu^j\right] &= L_{eu} + \frac{1}{2}\mu\left(Y_{ki} Y_{ki} - \Phi^i \Phi^i\right) \\ &+ \frac{1}{2}\mu\Lambda\left(\chi_{ki} \chi_{ki} - H_k^i H_k^i\right),\end{aligned} \quad (81)$$

其中除了熟悉的粘性系数 $\mu$，还有具有面积量纲的新物性参数 $\Lambda$，后者可以关联到流体元的特征尺度。从上述二次型拉氏量密度可以得到速度场和涡旋场的控制方程

$$\rho \frac{dV_i}{dt} - f_i + \partial_i p = \mu \nabla^2 V_i + \mu \varepsilon_{ilm}\left(2 A_k^l \partial_k V_m + V_m \partial_k A_k^l\right) \\ - \mu h_{ij,mn} V_j\left(A_k^m A_k^n + \Lambda B_k^m B_k^n\right), \quad (82)$$

$$\Phi^i = \Lambda\left(\partial_k H_k^i + \varepsilon_{ipq} A_k^p H_k^q\right), \quad (83)$$

$$\varepsilon_{ilm} V_l Y_{jm} = \Lambda\left(\partial_t H_j^i + \varepsilon_{ipq} \Phi^p H_j^q\right) \\ - \Lambda \varepsilon_{jlm}\left[\partial_l\left(\varepsilon_{ipq} V_p \chi_{mq}\right) + \varepsilon_{irs} \varepsilon_{spq} A_l^r V_p \chi_{mq}\right], \quad (84)$$

其中

$$h_{ij,mn} = \varepsilon_{kim} \varepsilon_{kjn} = \delta_{ij} \delta_{mn} - \delta_{in} \delta_{jm}。 \quad (85)$$

加上连续方程 $\partial_i V_i = 0$ 就是关于 16 个场量 $\{p, V_i, A_k^j, \Phi^j, i,j,k,l=1,2,3\}$ 的完整方程组。但是注意到守恒关系(80)现在变成

$$\partial_t \Phi^i = \varepsilon_{ilm} V_l D_k Y_{km}, \qquad (86)$$

似乎是一种不包含任何物性参数的运动转换关系：从主流搓出小尺度涡旋，而(84)式的左边项更是展现低维度的搓揉作用。

设单位矢量

$$\mathbf{n} = \cos\varphi\sin\theta\mathbf{e}_1 + \sin\varphi\sin\theta\mathbf{e}_2 + \cos\theta\mathbf{e}_3$$

表示速度的方向，$V$ 表示速率，$\mathbf{m}$ 表示 $\mathbf{n}\times\nabla V$ 的方向，或记为

$$\mathbf{m} = -\left(\cos\varphi\cos\theta\sin\phi + \sin\varphi\cos\phi\right)\mathbf{e}_1 + \\ \left(\cos\varphi\cos\phi - \sin\varphi\cos\theta\sin\phi\right)\mathbf{e}_2 + \sin\theta\sin\phi\mathbf{e}_3.$$

这样，以 $\mathbf{e}_1$ 表示滑移面方向，$\mathbf{e}_3$ 表示滑移方向，我们可以用旋转 $R(\varphi,\theta,\phi)$ 表示流体元的姿态标架，即取涡旋场的可积部分为

$$\mathbf{w}^1 = \sin\varphi d\theta - \cos\varphi\sin\theta d\phi,$$
$$\mathbf{w}^2 = -\sin\varphi\sin\theta d\phi - \cos\varphi d\theta,$$
$$\mathbf{w}^3 = -\cos\theta d\phi - d\varphi.$$

容易验证，用这个涡旋场导出的速度场强为

$$Y_i = DV_i = n_i dV,$$

从而保证了用姿态标架定义的涡旋场不会形成搓揉作用。由此不必动用控制方程求解的可积涡旋场可以用作涡旋场的规范条件。

要对流场方程组（82）-（84）（加上连续方程）进行求解，特别是数值求解，需要明确边界条件。这里提出如下边界条件：(1) 速度在固体边界上仍是无滑移。(2) 自旋场在固体边界上取零值。(3) 弯扭场用边界曲面的主曲率标架来标定，沿垂直于曲面方向标架取不变。由于弯扭场总是伴随速度场出现，在速度等于零的地方，弯扭场如何取值是无关紧要的；对于运动的固体壁面，则可以采认壁面附近是层流（无搓揉作用），如同前面定义的姿态标架，通过速度场来确定弯扭场。(4) 无回流稳态出口可取场量沿流向不变。(5) 单向层流入口处可取速度为层流分布、涡旋场等于零。

下面结合我们建立的新流动理论和观察阐释涡旋的形成机理。平直滑移流动（比如平板间的 Couette 流动，图 7-3）中流体分层序化滑移，各处流体元简单同构（弯扭场处处为零,可简单平移进行速度比较）；非平直滑移流动中的粘性力作用则有非零弯扭场与

速度场的耦合、容易触发小尺度涡旋。以雷诺数增加过程中圆柱绕流后面的涡旋的产生与演化为例，可以想象这样一个有待证实的机制：在柱面曲折滑移流动的局部始终存在一个欲以滚动代替滑动的紧张状态，这种紧张状态累积到一定的程度会突然以一个（一系列）耗散小涡的方式释放，即小尺度涡旋的触发具有量子化特性和一定的随机性，触发的小尺度涡旋有三种发展：（1）汇入或增强形成一个大涡；（2）通过集群效应改变或产生大尺度涡旋结构，从而影响速度场；（3）简单耗散掉。

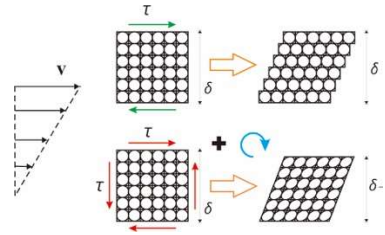

**图 7-3** Couette 流动的滑移和剪切理解
**Figure 7-3** Distinct understandings of slip and shear in the Couette flow.

## 7.3 与已有理论和思想方法的不同

下面讨论本节介绍的流动理论与各种已有理论和思想方法的不同。经典流动理论是一个宏观唯象理论，新流动理论也是一个宏观唯象理论。一种经常性的观点认为，满足连续性假设的宏观唯象理论只有唯一的架构，即现有的（分割聚合、构型映射）经典连续介质理论的架构，剩下可做的只能是去调整本构关系（对流体而言就是各种非牛顿流体模型）。已有的一些脱离经典的做法，往往是增加内变量（比如赋予质点以液晶指向、角速度乃至变形等）而不改变架构，其必然导致理论的不自洽（比如要定义微元体的转动惯量等）；本节则借助于微分几何这一现代数学工具，尝试建立连续介质理论的一种新的（微尺度覆盖、非同胚演化）框架，其中弹性变形和粘性流动具有不同的流形构造。在新理论中，湍流和层流的表达是一致一体的，我们**把湍流定义为涡旋动力学启动的流动（涡旋场方程不是零-零的平衡/有涡旋力矩/涡旋场强和自旋场不处处为零），湍流是搓出来的！**因此，本节给出的流动理论，既不是进行本构修订的非牛顿

流体模型，也不是简单扩展内变量空间的广义连续介质理论。

本节提出的流动理论应该不会出现在玻尔兹曼方程的截断上、也不会是不封闭的BBGKY层级理论的无穷无尽的场论逼近的某一环。实际上，对新理论的动力学方程展开分析，可以看到，复杂流动（湍流）中除了动量的输运，还有粘性氛围下多维度的力矩平衡，从主流（动量输运）中搓出的涡旋在动力学上引出了新的元素——力矩，相关的演化方程具有完整性和独立性，它们与雷诺平均导出的不封闭方程有本质的不同。

新流动理论包含以下三个基本观点：（1）作为流动最小分析单元（该部分流体的运动状态是时空点的函数）——流体元是开放的，流体是流体元的覆盖；（2）流体元除了宏观速度，还有时空内结构（开放系统的耗散结构），不可压缩流体没有疏密的变化，这个内结构就是分子尺度滑移导致的取向结构，时间方向引出自旋场，空间方向就是弯扭场；（3）采用最小置换与最小耦合原理构造流体系统的拉氏量密度，即建立宏观唯象的流动规范场论。

当流动为平直层流时，涡旋场处处为零，新理论和Navier-Stokes方程完全相同（方程（83）和（84）变成"零=零"，方程（82）的右边第二及以后各项等于零），但对于曲折流动，无论是否是湍流状态，新理论都将给出与N-S方程不同的预测。根据新理论及其对湍流的定义，湍流是通过粘性作用从局部"搓"出来的，粘性要求流体紧贴固体壁面滑移，除了拐点、角点这些流体取向突变处的强搓区域，在管流入口、边界层起点也都自然存在"搓"的机制，也就是说湍流的种子早已存在（微弱的小尺度的涡旋有时或难以察觉）。大尺度的涡旋必伴随涡核的存在，涡核区必有非零的弯扭场强，定常的大尺度涡旋是非常少见的，主流、大尺度涡旋和小尺度涡旋的相互作用、相互影响非常复杂（不仅是物理上的，三维旋转的不可交换性也会带来不同方向旋转的强耦合），如果细加分析，会发现强小尺度涡旋区域的猝发（指数增长）机制和流向涡自保持（强拟序、弱耗散）性质。简言之，在新理论中，速度场和涡旋场的非线性机制是大大丰富了，搓是从力到力矩（平流到涡流）的基本机制，由此用失稳解释层流转化为湍流的重要性也许会大大弱化。

# 8 结语

本文试图借对流动手性相关方面的讨论，比较系统地从基础流体力学到湍流理论进行既基础又前沿的介绍。所用的是现代几何拓扑的语言，包括现代场论的思想技术。

除了背景知识介绍，第五节和第六节的讨论主要是综述性质。第七节介绍的则是一个新的流动唯像规范场论[59]，试图对一般流动，尤其是湍流，提出新的观点、主张和模型。

限于篇幅，我们没有对辛（simplectic）流形和哈密顿场展开讨论。比如，第 5 节（21）式实际上是关于李-泊松结构的相应哈密顿方程的一个特例。后者对于研究余伴随轨道和可积性问题非常重要，这里Casimir 不变量标定不同的余伴随轨道，而对于第 6 节讨论的纽结，其同痕类对应于一条余伴随轨道，所以纽结分类问题就变成了卡希米（Casimir）不变量分类的子问题。我们也没有展开讨论可压缩流动和其它等离子体流体模型，它们中同样有全局螺度拓扑问题和局部螺旋度几何问题（见[14,63,4]），它们可通过引入半直积来对它们扩展第五节的描述，以及联系其中的同痕纽结和余伴随轨道。

描述高维流动时，微分几何的语言是无法避免的，除了提供统一的观点，技术上，对高维流动的理解对于研究低维被动标量的手性问题也是有用的，因为后者可视为前者的柱条件退化[64]。还有很多其它相关问题，比如相对论流体力学和量子流体，我们也不能全面讨论。

本文讨论的流动，包括第 7 节的湍流唯像方程，都假设解空间具有很好的微分结构。理想流动则只是一个特殊的同胚变换，由此我们介绍了比较丰富的现代结果。湍流研究如何从中直接受益是个有趣的课题。对于伽辽金截断系统，显然具有很好的微分结构，而且螺度还是守恒的，这在关于湍流的统计力学分析中非常重要[65,66,67,68,69]及本专题另文[70]。比如，文献[67]就伽辽金截断的无粘水动力学型方程中守恒的螺度之于拓扑流体动力学指出两方面：一方面，无粘但截断的方程继续保持螺度守恒，但是并不确信存在某虚拟速度使得涡冻结于其上，也就是并不保证涡管纽结的同痕变化（拓扑不变性）；另一方面，螺度的纽结拓扑图像并不受影响。换句话说，螺度拓扑

学内涵解释无关动力学，但是螺度的动力学守恒[16]并不表示拓扑不变性。果然，Moffatt[71]接着发表文章给出了伽辽金截断的 Euler 方程破坏拓扑不变性的具体例子。而 Hao, Xiong 和 Yang[16]则展示了在一些特殊的非理想流动中涡冻结的经典虚拟速度是存在的。对于湍流而言，目前 Navier-Stokes 方程解的存在唯一性还是悬而未决的"世纪问题"。而现有的理论和证据表明充分发展湍流很可能不具备那些很好的微分结构，甚至需要分形的语言来描述，更勿论经典的虚拟冻结。但是，对于湍流如果能发展某种"广义"的虚拟冻结理论和技术，其"广义"的内涵明确且可实现，比如是某种统计性质明确的随机场，那么这对于复杂湍流的结构控制是非常有潜力的，因为它意味着理想流动结果可以适当地"移植"以用于湍流研究。

根据上一段讨论，对于湍流这样的复杂流动，传统拓扑流体力学严格来说是不存在的，但是某种广义的的统计拓扑流体力学是可能的，而且使得经典拓扑流体力学框架下的结果发挥重要作用。第 6 节讨论的螺度纽结拓扑图象其实可以提供我们关于日常绳结的直觉：比如 Moffatt[72]提出了理由说明流体系统弛豫过程中能量下降需伴以纽结通量管的长胖，当通量管长胖到互相接触后，由于拓扑不变性（障碍），就不能再长了，故而能量有个下界[73]。螺度拓扑效应统计理论分析[69]暗示螺度可能束缚流动压缩模，最近确实发现对于经典流体和（玻色-爱因斯坦凝聚）量子流体都是有效的[74,70]，而且对力和热问题同样具有通过螺度来有效控制的很大可能性。这在一定程度上也可通过这种涡绳纽结的螺度界定能量弛豫机制来理解。给定能量时，螺度（绝对值）越大，弛豫的余地就越小，反而言之，涡绳将流体"捆"得越紧[70]。

上面提到螺度束缚流动压缩模，这可以体现为气动声学降噪[69,70,74]，力热声是耦合的，也是流体力学工程问题中的核心。确实，已经有研究发现旋转流动中的螺度减阻效果[75]。旋转模态从物理粒子到星球，黑洞，星系都是普遍而根本的，理解旋转的意义和效果当然是重要的。经典中性流体中螺度效应似乎在旋转流动中更有可能有的明显体现，也是基础研究的热点。比如，Pouquet 和 Mininni[76]研究了的旋转和螺度的相互影响；Linden-Bell 和 Katz[77]讨论了旋转流动的 Lagrange 变分原理；后者建议的特殊标记后来被 Yahalom[78]用于提出新的对称群，并如第六节谈到的，通过 Noether 定理证明其产生螺度守恒。由此，我们看到螺度对于流体力学的应用和基础的联系是如此的紧密和直接！

---

[16] 螺度守恒系统还可以是时间可逆但是相体积不守恒，也即不满足 Liouville 定理的系统，所以系统不会像截断的 Euler 一样趋于统计平衡。对于由此生成的非平衡动力学系综人们可以进行一些平衡与非平衡统计计算，揭示湍流基础而又根本的性质。这方面的讨论可见于本专刊另文[61]。

# Geometrical and topological description of chirality-relevant flow structures

Wennan Zou[1], Jian-Zhou Zhu[2*], and Xin Liu[3]

[1] *Institute for Advanced Study, Nanchang University, Nanchang 330031, China;*
[2] *Su-Cheng Centre for Fundamental and Interdisciplinary Sciences, Gaochun, Nanjing, 211316, China;*
[3] *Institute of Theoretical Physics, Beijing University of Technology, Beijing 100124, China.*

Issues relevant to the flow chirality and structure are focused, while the new theoretical results, including even a distinctive theory, are introduced. However, it is hope that the presentation, with a low starting point but a steep rise, is appropriate for a broader spectrum of audiences ranging from students to researchers, thus illustrations of differential forms and relevant basic topological concepts are also offered, followed by the demonstration with formulation of differential forms of the classical Navier-Stokes flow theory and the discussions of recent studies in fundamental fluid mechanics and turbulence.

**keyword 1, keyword 2, keyword 3**

**PACS:** 47.27.-i, 47.27.Eq, 47.27.Nz, 47.40.Ki, 47.85.Gj